\documentstyle[12pt,amstex,epsf,righttag]{article}
\newcommand{\single}{}

\newcommand{\la}{\lambda}
\newcommand{\La}{\Lambda}

\newcommand{\Ga}{\Gamma}
\newcommand{\ep}{\epsilon}

\newcommand{\g}{\varphi}
\newcommand{\vecv}{{\boldmath $v$}}

\newcommand{\vece}{{\boldmath $e$}}
\newcommand{\vecr}{{\boldmath $r$}}
\newcommand{\vecm}{{\boldmath $m$}}
\newcommand{\veca}{{\boldmath $a$}}

\newcommand{\vecc}{{\boldmath $c$}}

\newcommand{\vecf}{{\boldmath $f$}}

\newcommand{\vecn}{{\boldmath $n$}}

\newcommand{\vecp}{{\boldmath $p$}}

\newcommand{\vecw}{{\boldmath $w$}}

\newcommand{\vecbeta}{{\boldmath $\beta$}}

\newcommand{\vecep}{{\boldmath $\ep$}}

\newcommand{\va}{\text{\veca}}

\newcommand{\vc}{\text{\vecc}}

\newcommand{\ve}{\text{\vece}}
\newcommand{\vf}{\text{\vecf}}

\newcommand{\vm}{\text{\vecm}}
\newcommand{\vn}{\text{\vecn}}

\newcommand{\vp}{\text{\vecp}}

\newcommand{\vr}{\text{\vecr}}

\newcommand{\vv}{\text{\vecv}}
\newcommand{\vw}{\text{\vecw}}

\newcommand{\vbeta}{\text{\vecbeta}}

\newcommand{\vep}{\text{\vecep}}

\newcommand{\C}{{\Bbb C}}

\newcommand{\Z}{{\Bbb Z}}
\newcommand{\Q}{{\Bbb Q}}
\newcommand{\bP}{{\Bbb P}}
\newcommand{\N}{{\Bbb N}}
\newcommand{\kahler}{{\Bbb K}}
\newcommand{\Mod}{{\cal M}(\vr)}
\newcommand{\GIT}{/\negthickspace/}
\newcommand{\CY}{{\C}^d/\Ga}  

\newcommand{\facetset}{\La}
\newcommand{\athfacet}{{\cal F}_{a}}
\newcommand{\bthfacet}{{\cal F}_{b}}

\newcommand{\config}{configuration}
\newcommand{\Nb}{\overline{N}}
\newcommand{\Mb}{\overline{M}}
\newcommand{\vwb}{\overline{\vw}}
\newcommand{\vvb}{\overline{\vv}}

\newcommand{\A}{{\cal A}}  
\newcommand{\ct}{({\C}^{*})}
\newcommand{\Ka}{K\"ahler\ }
\newcommand{\what}{\widehat}
\newcommand{\Qr}{Q(\what{\vr})}
\newcommand{\amu}{a_{\mu}}
         
\newcommand{\tM}{\widetilde{M}}
\newcommand{\ra}{\rightarrow}
\newcommand{\piQ}{\pi_{\Q}}
\newcommand{\Mtot}{M^{(0)}}
\newcommand{\Msub}{M^{(1)}}
\newcommand{\cone}{C_{\text{basic}}}
\newcommand{\pos}{\mathop{\mathrm{cone}}}
\newcommand{\conv}{\mathop{\mathrm{conv}}}
\newcommand{\rec}{\mathop{\mathrm{rec}}}
\newcommand{\Spec}{\mathop{\mathrm{Spec}}}
\newcommand{\Proj}{\mathop{\mathrm{Proj}}}
\newcommand{\Ker}{\mathop{\mathrm{Ker}}}
\newcommand{\Hom}{\mathop{\mathrm{Hom}}}
\newcommand{\Aut}{\mathop{\mathrm{Aut}}}
\newcommand{\codim}{\mathop{\mathrm{codim}}}
\newcommand{\Sym}{\mathop{\mathrm{Sym}}}
\newcommand{\hilb}{\mathrm{Hilb}}  
\newcommand{\Vect}{{\Bbb W}}
\newcommand{\skima}{\thinspace}
\newcommand{\Hilb}{\hilb^{\Ga}({\C}^{d})}
\newcommand{\Hn}{\hilb^n({\C}^d)}  
\newcommand{\susy}{supersymmetry}
\newcommand{\kahilb}{r}
\numberwithin{equation}{section}   

\begin{document}
\begin{center}
{\large
{\it Preliminary Version, \ June 1998}}
\end{center}
%
\vspace{2cm}

\begin{center}
{\LARGE {\bf  \Ka Moduli Space for a D-Brane 
\vspace{0.5cm}

at Orbifold Singularities}}
\vspace{0.5cm}

\begin{center}
{\large {\boldmath $Dedicated$ $to$ $the$ $memory$ $of$ $a$ $cat$}}
\end{center}
\vspace{0.7cm}

{\large Kenji  \ Mohri}
\vspace{0.5cm}

{\it Theory Group, Institute of Particle and Nuclear Studies},

{\it High Energy Accelerator Research Organization (KEK)},

{\it Oho~1-1 Tsukuba, Ibaraki 305-0801, Japan}

mohri@@theory.kek.jp



\end{center}
\vspace{4cm}

\begin{abstract}
We develop a method to analyze systematically 
the configuration space of a D-brane 
localized at the orbifold singular point 
of a Calabi--Yau $d$-fold of the form ${\Bbb C}^d/\Gamma$
using the theory of toric quotients.
This approach elucidates 
the structure of the K\"ahler moduli space associated with
the problem. 
As an application, we compute the toric data of 
the $\Gamma$-Hilbert scheme.
\end{abstract}
\vfill
\begin{flushright}
hep-th/9806052
\end{flushright}
\single

\section{Introduction}

The \config\ space of a D-brane localized 
at the orbifold singularity of a Calabi--Yau $d$-fold
of the form $\CY$, 
where $\Ga$ is a finite subgroup of $\text{SU}(d)$, 
is an interesting object to study,
because it represents the ultra-short distance geometry
felt by the D-brane probe \cite{DKPS}, 
which may be different from the geometry of bulk string.
On the mathematical side,
the D-brane \config\ space corresponds
to a generalization 
of the Kronheimer construction of 
the ADE type hyper-\Ka manifolds \cite{kronheimer}
to higher dimensions, which has been studied 
by Sardo Infirri \cite{sardo,infirri}. 
He has shown that the D-brane \config\ space 
is a blow-up of the orbifold $\CY$, 
the topology of which depends 
on the \Ka (or Fayet--Iliopoulos) moduli parameters~;
Moreover he has conjectured that for $d=3$,
the D-brane \config\ space is a smooth Calabi--Yau
three-fold for a generic choice of the \Ka moduli parameters.
The case in which $\Ga$ is Abelian is 
of particular importance,
because then the \config\ space is a toric variety,
which enables us to employ various methods 
of toric geometry to study it.
Using toric geometry,
several aspects of the D-brane \config\ space 
have been studied so far
\cite{DG,DGM,DM,greene,mohri,MR,muto,ray}.

Our aim in this article is to give a method 
to analyze systematically 
the structure of the \Ka moduli space associated with
the D-brane \config\ space
which releases one from the previous brute force calculations, 
for example see \cite[(53--74)]{mohri}. 
It turns out that the theory of toric quotients 
developed by Thaddeus \cite{thaddeus} provides us with 
the most powerful tool to investigate 
the D-brane \config\ space.
This approach has already been taken in \cite{infirri},
where the analysis of the toric data
is reduced to the network flow problem  
on the McKay quiver defined by the orbifold.

To save the notation, we consider
only cyclic groups for $\Ga$, 
but the generalization to an arbitrary Abelian group,
that is, a product of several cyclic groups,
should be straightforward.
\vspace{0.5cm}

The organization of this article is as follows~:

In section 2, we explain in detail 
the construction by Thaddeus \cite{thaddeus}
of quasi-projective toric varieties and 
their quotients by subtori 
in terms of rational convex polyhedra.
This formulation gives us a clear picture of 
the \Ka moduli space 
associated with a toric quotient \cite{KSZ,thaddeus}.

In section 3, we describe the \config\ space of a D-brane 
localized at the orbifold singularity as a toric variety
obtained by a toric quotient of an affine variety
closely following the treatment 
by Sardo Infirri \cite{infirri}.
Then we give typical examples
of phases of the D-brane \config\ spaces
for Calabi--Yau four-fold models. 

Section 4 is devoted to 
an application of our construction
of the D-brane \config\ space 
to the $\Ga$-Hilbert scheme 
\cite{ito-nakajima,ito-nakamura,nakajima,nakamura,reid},
which is roughly the moduli space of $|\Ga|$ points
on $\C^d$ invariant under the action of $\Ga$,
in the hope that the investigation 
of various Hilbert schemes
sheds light on the geometrical aspect of D-branes 
on Calabi--Yau varieties \cite{BBMOOY,BVS,OOY}. 
\vspace{0.5cm} 

For textbooks or monographs 
dealing with various aspects of toric varieties 
and related topics,
consult \cite{AGV,cox,ewald,fulton,GKZ,oda,sturmfels,ziegler},
as well as the physics articles \cite{AGM,MP,witten1}, 
which contain introductory materials intended for physicists.

\single
\section{Toric Varieties and Its Quotients}
\subsection{Polyhedra and Quasi-Projective Varieties}
Let $N$ be a lattice of rank $p$ 
and $M=N^{*}$ be the dual lattice.
Let 
$T=\Hom(M,{\C}^*)
\cong N\otimes_{\Z}{\C}^*
\cong \ct^{p}$
be the associated torus.
Then we have the following identification~:
\begin{align}
M&=\Hom(T,{\C}^*),
\qquad \text{characters of } T,\\
\label{character}
N&=\Hom({\C}^*,T), 
\qquad \text{1-parameter subgroups of } T.
\label{1-parameter}
\end{align} 
Let $P$ be a $p$-dimensional convex polyhedron 
in the vector space $M_{\Q}$.
We want to associate a quasi-projective toric variety
to the data $(M,P)$,
which we denote by $X(M,P)$ or simply by $X(P)$
if no confusion occurs.

$P$ can be represented 
as an intersection of half-spaces as follows~:
\begin{equation}
P=\left\{
\vm\in M_{\Q}\left|\ 
\langle \vm,\vv_a \rangle \geq t_a,\
\forall a\in \facetset 
\right.
\right\},
\label{convex-polyhedron}
\end{equation}
where ${\vv_a}\in N$ and  ${t_a}\in \Q$ and
$\facetset$ is an index set.

For technical reason, 
we put the following assumptions on $P$~:
\begin{enumerate}
\item
Each $\vv_a$ is a {\it primitive} vector, that is,
for any integer $n>1$,
$(1/n)\skima \vv_a \not\in N$.
\item The expression of $P$ (\ref{convex-polyhedron})
is reduced in the sense that
the omission of the $a$~th inequality 
in (\ref{convex-polyhedron}) gives rise 
to a polyhedron strictly larger than $P$
for any $a\in \La$.
\item The vector space defined by  
$\{\vm\in M_{\Q}|\ \langle \vm,\vv_a\rangle=0,
\ \forall a\in \facetset\}$,
which is the maximal vector subspace in $P$,
is equal to  $\{\bold{0}\}$.
\end{enumerate}

The $a$~th facet of $P$, 
which we denote by $\athfacet$,
is given by
\begin{equation}
\athfacet:=
\left\{
\vm\in P\left|\ 
\langle \vm,\vv_a \rangle=t_a
\right.
\right\},         
\end{equation}
which shows that ${\vv_a}$ is an inner normal vector 
to $P$ at $\athfacet$.

Here let us describe combinatorics of $P$ 
\cite[Lecture 2.2]{ziegler}.

By the {\it face lattice} of $P$, we mean 
the set of all the faces of $P$ 
partially ordered by inclusion
relation, which is denoted by $L(P)$. 
We also denote the proper part of it by
$\overline{L}(P):=L(P)\backslash (\emptyset, P)$. 
For each $F\in \overline{L}(P)$,
we define a subset $I(F)$ of $\facetset$ by
\begin{equation}
I(F):=\{a\in \facetset
|\ F\subset \athfacet \},
\end{equation}
where $\text{card}\ I(F)\geq \codim F$. 
Then each $F\in \overline{L}(P)$ can be represented as
an intersection of facets as follows~:
\begin{equation}
F=\bigcap_{a\in I(F)}\athfacet.
\label{intersection}
\end{equation}
It is also convenient to set formally
$I(P)=\emptyset$,
$I(\emptyset)=\facetset$
and to regard (\ref{intersection})
valid even for $F=\emptyset, P$. 
Then the intersection $\cap$ of any two elements of 
$L(P)$ can be described in an obvious manner, that is,
\begin{equation}
F_1\cap F_2=\bigcap_{a\in I(F_1)\, \cup \, I(F_2)}
\athfacet.
\end{equation}
Again for $F_1,F_2\in L(P)$,
let $F_1\cup F_2\in L(P)$ be the smallest 
among those which contains both $F_1$ and $F_2$.
The operation $\cup$ is called {\it join}.
We see that for $F_1,F_2\in L(P)$,
\begin{equation}
F_1\cup F_2=\bigcap_{a\in I(F_1)\,\cap \, I(F_2)}
\athfacet.
\end{equation}


Define a rank $(q+1)$ lattice by 
$\tM:=\Z\times M$ 
and define a cone $C(P)$
in $\tM_{\Q}$,
which is called the homogenization of $P$
\cite[Lecture 1.5]{ziegler}, by
\begin{align}
C(P)&=\text{closure of }
\left\{ \la (1,\vm) \ \left| \ \la\in {\Q}_{\geq 0},
\ \vm\in P  \right.\right\}
\text{ in } \tM_{\Q}, \nonumber \\ 
&=\left\{ \la (1,\vm)\ \left|\  \la \in {\Q}_{ > 0},
\ \vm \in P \right.
 \right\}
+\{ \bold{0} \}\times \rec P,
\end{align}
where a Minkowski sum is used in the second line
and $\rec P$ is 
the {\it recession cone} of $P$ defined by
\begin{equation}
\rec P=\left\{\  \vm\in M_{\Q} 
\left|\ \vm^{'}+\la \vm\in P,\ 
\forall \vm^{'}\in P, \ 
\forall \la \in {\Q}_{>0} \right.\right\}.
\end{equation}
In our case a more concrete expression is possible~:
\begin{equation}
\rec P
\cong
\left\{
\vm\in M_{\Q}
\left|\ 
\langle \vm,\vv_a\rangle \geq 0, 
\ \forall a\in \facetset
\right.
\right\}.
\label{alternate}
\end{equation}
%
$C(P)\cap \tM$ has a structure of 
a graded $\rec P$-algebra
graded by its first component, that is,
$(C(P)\cap \tM)_k:=C(P)\cap (\{k\}\times M)$ and
$(C(P)\cap \tM)_0=\rec P$,
which leads us to the following definition of 
$X(P)$ as a quasi-projective variety which is 
projective over an affine variety 
\cite[(2.9)]{thaddeus}
\begin{equation}
X(P):=\Proj \left(C(P)\cap \tM\right)
{\longrightarrow}\ 
X_0(P):=\Spec \left(\rec P\right).
\label{Proj}
\end{equation}
Strictly speaking, every scheme $X$ in this article, 
either affine or projective,
should be replaced by the set of 
its $\C$-valued points 
$X(\C):=\Hom_{\C}(\Spec \C, X)$ 
\cite{oda}.

To be more explicit, we construct $X(P)$
by the following procedure.
First let $(k_1,\vm_1)$, \dots , $(k_s,\vm_s)$
be the generators of 
$C(P)\cap \tM$.
Then we have an embedding 
of $X(P)$ in the weighted projective space
$\bP(k_1,\dots ,k_s)$, where a degree $k_j$ may be 0~;
more precisely, the degree zero generators 
of $C(P)\cap \tM$ are those of $\rec P\cap M$.
The ambient space $\bP(k_1,\dots ,k_s)$ of $X(P)$ 
admits a following symplectic quotient realization~:
\begin{equation}
\bP(k_1,\dots ,k_s)=
\left.
\left\{ (z_1,\dots ,z_s)\in  \C^s
\left|
\  \sum_{j=1}^s k_j\ |z_j|^2=1
\right.
\right\}
\right/
\text{U}(1).
\label{ambient}
\end{equation} 

Second let $\psi$ be the lattice surjection {}from
$\Z^s$ to $C(P)\cap \tM$ defined by
\begin{equation}
\psi(\vc):=\sum_{j=1}^s  c_j (k_j,\vm_j).
\end{equation}
Then $\Ker \psi$ is the lattice that represents
the relations between the generators of $C(P)\cap \tM$.
We convert them 
to equations for the homogeneous coordinates
$(z_j)$ of $\bP(k_1,\dots ,k_s)$,
which is called the F-flatness equations in physics terminology~:
\begin{equation}
\prod_{c_j>0}
 z_j^{c_j}
=\prod_{c_j<0}
z_j^{-c_j},
\quad \vc\in \Ker \psi,
\label{F-flat}
\end{equation}
where the degree of $z_j$ is $k_j$.

We now get a symplectic quotient realization of $X(P)$~:
\begin{equation}
X(M,P):=
\left\{
(z_j) \in \C^s 
\left.
\left|
\begin{gathered}
\sum_{j=1}^s k_j\ |z_j|^2=1 \\ 
\text{F-flatness equations}\  (\ref{F-flat})
\end{gathered}
\right\}
\right.
\right/ \text{U}(1).
\label{X(P)}
\end{equation}

If $P$ itself is a polyhedral cone in $M_{\Q}$,
then  $C(P)\cong {\Q}\times P$ so that
$\Proj \left(C(P)\cap \tM\right)$ is isomorphic to 
$\Spec \left(P\cap M\right)$, that is, 
$X(P)$ is an affine variety.

Another extreme case is when $P$ is a bounded polyhedron, 
that is, polytope. Then $X(P)$ is a projective variety.
\begin{flushleft}
{\it Example.}
Let $M=\Z^2$ and 
$P=\pos \{\skima\bold{0},\ (1/2) \ve_1, \ (1/3) \ve_2\}
\subset M_{\Q}$.
Then $C(P)\cap \tM$ is freely generated by 
$(1,\bold{0})$, $(2,\ve_1)$ and $(3,\ve_2)$,
so that $X(P)=\bP(1,2,3)$.
\end{flushleft}
\begin{flushleft}
{\it Example.}
Let $M=\Z^2$ and 
$P=\conv\skima \{\skima 3\ve_1,\ \ve_1+\ve_2,\  3\ve_2\}
+\pos \{\skima\ve_1, \ \ve_2\}$.
Then
\begin{equation}
X(P)=
\left\{
\left(x_1,x_2;T_1,T_2,T_3\right)\in \C^2\times \bP^2
\left|\
\begin{gathered}
x_1T_3-x_2^2T_2=0,\ \ 
x_2T_1-x_1^2T_2=0 \\
T_1T_3-x_1x_2T_2^2=0   
\end{gathered}
\right.
\right\},           \nonumber
\end{equation}
which is projective over the affine variety
$X(\rec  P)=\C^2$.
\end{flushleft}
The $T$-action on the homogeneous coordinates is given by
\begin{equation}
z_j\ra \la^{\langle \pmb{m}_j, \pmb{n}\rangle}z_j,
\quad \vn\in N,\ \la\in \C^*,
\label{homogeneous}
\end{equation} 
where we regard $\vn\in N$ as a 1-parameter subgroup
of $T$ according to (\ref{1-parameter}).
In an evident way, (\ref{homogeneous})
induces a $T$-action on $C(P)\cap \tM$, 
which defines a linearization, that is, a lifting to
an ample line bundle, 
of the $T$-action on the base $X(P)$.
\subsection{Toric Varieties {}from Fans}
Now that we have given a variety $X(M,P)$ associated with 
a polyhedron $P\subset M_{\Q}$, 
it is natural to ask for the fan 
in $N_{\Q}$ that yields $X(M,P)$ as a toric variety.

To describe the fan associated with $X(M,P)$,
let us first define the following function~:
\begin{equation}
h(\vn):=\text{min}\left\{ \langle \vm^{'}, \vn\rangle :\  
 \vm^{'}\in P \right\},
\label{def-h}
\end{equation}
which is called the {\it support function} 
of $P\subset M_{\Q}$ 
\cite[Appendix]{oda}.
Note that the domain of definition of $h$, 
which we denote by $\text{dom }h$, is 
\begin{equation}
\text{dom }h
=\pos 
\left\{\skima
\vv_a\left|
\ a\in \facetset
\right\}
\right.
\subset N_{\Q},
\end{equation}
which is $p$ dimensional owing to the third assumption
on $P$ that we put earlier.

Now define a cone $C(F)$ in $N_{\Q}$ for 
$F\in L(P)\backslash \emptyset$ by
\begin{equation}
C(F):=\left\{\  
\vn\in \text{dom }h 
\left|\ 
\right.
 \langle \vm,\vn\rangle = h(\vn),
\ \ \forall \vm\in F 
\right\} \subset N_{\Q},
\end{equation}
which is call the {\it normal cone} of $F$.

To be more explicit, for the $a$~th facet of $P$,
$C(\athfacet)
=\pos\{\skima\vv_a\}={\Q}_{\geq 0}\skima \vv_a$ 
and for a lower dimensional face $F$,
\begin{equation}
C(F)=\pos \{\skima\vv_a |\ a\in I(F)\}
=\bigoplus_{a\in I(F)}\ {\Q}_{\geq 0}\skima \vv_a.
\end{equation}
We also  see that $C(P)=\{ \bold{0} \} \in N_{\Q}$ 
because we always assume that $\dim P=p$. 

Note that $\dim F+\dim C(F)=p$
and $F\in L(P)\backslash\emptyset$ can be recovered 
from $C(F)$ by
$$
F=\left\{
\vm\in P
\left|\ 
\langle \vm,\vn\rangle=h(\vn),\  \forall \vn\in C(F)
\right\}
\right..
$$

Moreover for $F_1, F_2\in L(P)\backslash \emptyset$, 
$C(F_1)$ is a face of $C(F_2)$ if and only if 
$F_2$ is a face of $F_1$,
and $C(F_1\cup F_2)=C(F_1)\cap C(F_2)$ is a common face of
$C(F_1)$ and $C(F_2)$. 
Thus we can define a fan in $N_{\Q}$ by
\begin{equation}
{\cal N}(P):=
\left\{
 C(F) 
\left|\  
F\in 
{L(P)\backslash \emptyset}
\right.
\right\},
\end{equation}
which we call the {\it normal fan} of $P$, and
the support of which is $\text{dom }h$.

We denote by $X^*(N,{\cal N}(P))$ 
the toric variety associated with
the data $(N,{\cal N}(P))$.
By definition,
$X^*(N,{\cal N}(P))$ has the following affine open covering~:
\begin{equation}
X^*(N,{\cal N}(P))
=\bigcup_{F\in L(P)\backslash \emptyset}
X(M,C(F)^{*}),            
\label{affine-open-cover}
\end{equation}
where 
$X(M,C(F)^{*})
=\Spec \left( M\cap C(F)^{*}\right)$, 
and for a cone $C\subset N_{\Q}$, its dual cone  
$C^{*}\subset M_{\Q}$ is defined by
\begin{equation}
C^{*}:=
\left\{
\vm\in M_{\Q}\left|\ 
\langle \vm,\vn\rangle\geq 0, \ \vn\in C
\right\}
\right..
\end{equation}

\begin{flushleft}
{\it Proposition 2.1.}
{\sl
$X^*(N,{\cal N}(P))$ is isomorphic to $X(M,P)$.}
\end{flushleft}

This follows 
{}from the fact that the affine open covering of 
$X^*(N,{\cal N}(P))$ described  in (\ref{affine-open-cover}) is
identical with that of $X(M,P)$
given in \cite[Proposition (2.17)]{thaddeus}.

The shape and the size of the polyhedron $P$ carry 
informations about the \Ka moduli parameters of $X(M,P)$, 
which are lost in converting $P$ into its normal fan 
${\cal N}(P)$.
Two polyhedra $P_1$, $P_2$ in $M_{\Q}$ 
are said to be {\it normally equivalent} 
if their normal fans  are isomorphic to each other,
that is, ${\cal N}(P_1)\cong {\cal N}(P_2)$.
\begin{flushleft}
{\it Example.} Let us take $M=\Z^2$ and a pair of
normally equivalent polyhedra
\begin{align}
P_1&=\conv \skima
\{\bold{0},\ \ve_1,\ \ve_2,\ \ve_1+\ve_2 \},\nonumber \\
P_2&=\conv\skima
\{\bold{0},\ 4\ve_1,\ 3\ve_2,\ 4\ve_1+3\ve_2 \}.\nonumber
\end{align}
Both $X(M,P_1)\subset \bP^3$ and $X(M,P_2)\subset \bP^{19}$ 
are isomorphic to
$\bP^1\times \bP^1$~; 
the \Ka moduli of the former and the latter are
$(1,1)$ and $(4,3)$ respectively.
\end{flushleft}
The use of the normal fan, however, is a far more  
efficient way to obtain the toric variety $X(M,P)$.


\subsection{Toric Quotient}
Let $P\subset M_{\Q}$ be a polyhedron,
and  $X(M,P)$ be 
the associated  quasi-projective variety.
Suppose that there is an exact sequence of lattices
\begin{equation}
0 \ra N^{'}\overset{\pi^*}{\ra} N 
\overset{i^*}{\ra} \Nb\ra 0, 
\label{exact-N}
\end{equation}
where $\text{rank}\  N^{'}=p-q$ and $\text{rank}\ 
 \Nb=q$,
then the dual sequence is also exact~:
\begin{equation}
0 \ra \Mb\overset{i}{\ra} M 
\overset{\pi}{\ra} M^{'}\ra 0.
\label{exact-M}
\end{equation}
A sublattice $N^{'}\subset N$ defines 
a subtorus
$T^{'}=N^{'}\otimes\C^*=\Hom(M^{'},\C^*)$ 
of rank $p-q$, which acts on $X(M,P)$.

Now we want to define 
the geometric invariant theory (GIT)
quotient of $X(M,P)$ by the action of $T^{'}$.

The graded ring $C(P)\cap \tM$ admits a natural $T^{'}$-action
and the $T^{'}$-invariant part
$\left( C(P)\cap \tM \right)^{T^{'}}$ is also a graded ring.
Then we define the quotient variety by 
\begin{equation}
X(M,P)\GIT T^{'}:=\Proj \left( C(P)\cap \tM \right)^{T^{'}},
\end{equation}
which is again projective 
over the affine variety defined by
the affine GIT quotient 
\begin{equation}
X_0(M,P)\GIT T^{'}:=\Spec \left(\rec P\cap M\right)^{T^{'}},
\end{equation}
where $(\rec P\cap M)^{T^{'}}$ is the degree zero part
of $\left( C(P)\cap \tM \right)^{T^{'}}$.

We immediately see 
that the GIT quotient variety admits 
a following toric realization~:
\begin{equation}
X(M,P)\GIT T^{'}
=X\left(\Mb, P\cap \pi_{\Q}^{-1} (\bold{0})\right),
\label{GIT=0}
\end{equation}
where $\Mb= \Ker \pi=M\cap \pi_{\Q}^{-1}(\bold{0})$ is 
the sublattice of $M$ fixed by $T^{'}$.

The corresponding symplectic quotient construction 
can be done as follows~:
In addition to the D-flatness equation in (\ref{ambient}) 
\begin{equation}
\sum_{j=1}^s k_j\ |z_j|^2=1
\label{1-ambient}
\end{equation}
for the ambient space $\bP(k_1,\dots ,k_s)$,
we put $p-q$ D-flatness equations associated with
$T^{'}$-action on $(z_j)$ with the \Ka 
(or Fayet--Iliopoulos) parameters  
$\vr=\bold{0}\in M^{'}_{\Q}$
followed by quotienting by $\text{U}(1)^{p-q}$.
More concretely, 
let $\vn^{'}_1,\dots ,\vn^{'}_{p-q}$ 
be the generators of $N^{'}$, 
each of which corresponds to a 1-parameter subgroup of 
$T^{'}\cong (\C^*)^{p-q}$.
Then the additional $p-q$ D-flatness equations 
can be written as
\begin{equation}
\sum_{j=1}^s {\langle \pi(\vm_j),\vn^{'}_l \rangle}\ 
|z_j|^2=0, \qquad l=1,\dots ,p-q.
\label{additional}
\end{equation}
A useful abbreviation of (\ref{additional}) is
\begin{equation}
\sum_{j=1}^s \pi(\vm_j)\ |z_j|^2 =\bold{0},
\label{0-vector}
\end{equation}
where we say that $z_j$ has $T^{'}$-charge $\pi(\vm_j)$.

Now we want to consider the toric quotient of $X(P)$ by $T^{'}$ 
with a nonzero \Ka moduli parameters
$\vr\in M^{'}_{\Q}$.
To this end let us take $\what{\vr}\in M_{\Q}$ such that
$\pi_{\Q}(\what{\vr})=\vr$ and consider the shifted polyhedron 
$P-\what{\vr}\subset M_{\Q}$.
The original generators $(k_j,\vm_j)$ of 
$C(P)\cap \tM$ are now shifted to
$(k_j,\vm_j-k_j \what{\vr})$ so that 
the $T^{'}$-charge of $z_j$ becomes
$(\pi(\vm_j)-k_j \vr)$.
This $T^{'}$ charge assignment 
for $(z_j)$ defines a new action of
$T^{'}$ on $(C(P)\cap \tM)$ 
which we denote by $T^{'}(\vr)$.
Then we can define 
the GIT quotient of $X(M,P)$ by $T^{'}(\vr)$ as
\begin{equation}
X(M,P)\GIT T^{'}(\vr):=\Proj  
\left(C(P)\cap \tM\right)^{T^{'}(\pmb{r})}
\label{quotient-def}
\end{equation}
which is also projective over the affine variety
\begin{equation}
X_0(M,P)\GIT T^{'}(\vr):=
\Spec \left(\rec P\cap M\right)^{T^{'}(\pmb{r})}.
\label{quotient-def-affine}
\end{equation}
The ambiguity in the choice of $\what{\vr}$, which 
is isomorphic to $\Mb_{\Q}$, does not
affect the definitions (\ref{quotient-def}),
(\ref{quotient-def-affine}).
In fact it only affects 
the $\overline{T}:=T/T^{'}$-linearization
of the quotient variety, which is irrelevant to us.

To see that the definition of 
$X(M,P){\GIT} T^{'}(\vr)$ above corresponds to
the change of the \Ka parameters to $\vr\in M^{'}_{\Q}$,
we have only to describe the corresponding
symplectic quotient construction of $X(M,P)\GIT T^{'}(\vr)$.
The D-flatness equations associated with $T^{'}(\vr)$ are 
\begin{equation}
\sum_{j=1}^s (\pi(\vm_j)-k_j \vr)\ |z_j|^2 =\bold{0}.
\label{modified}
\end{equation}
Combining (\ref{1-ambient}) and (\ref{modified}), 
we obtain the D-flatness equations associated with $T^{'}$
with the \Ka moduli parameters $\vr$~:
\begin{equation}
\sum_{j=1}^s \pi(\vm_j)\ |z_j|^2 =\vr.
\end{equation}
Thus we get $X(M,P)\GIT T^{'}(\vr)$ by
the following symplectic quotient of 
$X(P)$ by the $\text{U}(1)^{p-q}$-action 
with $\vr$ as a \Ka parameters~:
\begin{align}
X(M,P)\GIT T^{'}(\vr) &\cong
\left.
\left\{
\left[(z_j)\right]\in X(M,P)
\left|\
\sum_{j=1}^s \pi(\vm_j)\ |z_j|^2=\vr
\right.
\right\}\right/\text{U}(1)^{p-q}  \\
&\cong
\left\{
(z_j) \in \C^s 
\left.
\left|
\begin{gathered}
\sum_{j=1}^s k_j\ |z_j|^2=1,\ 
\sum_{j=1}^s \pi(\vm_j)\ |z_j|^2 =\vr\\
\text{F-flatness equations}\ (\ref{F-flat})
\end{gathered}
\right\}
\right.
\right/ \text{U}(1)^{p-q+1}.\nonumber
\end{align}

In the following we argue that the GIT quotient 
$X(M,P){\GIT} T^{'}(\vr)$ defined above can be realized as a
quasi-projective toric variety~:

We will show that $X(M,P){\GIT} T^{'}(\vr)$ 
can be realized as a 
quasi-projective toric variety generalizing (\ref{GIT=0})~: 

\begin{flushleft}
{\it Proposition 2.2.}
{\sl Fix $\what{\vr}\in M_{\Q}$ such that 
$\pi_{\Q}(\what{\vr})=\vr$ for $\vr\in {M}^{'}_{\Q}$, and
let $\Qr\subset \Mb_{\Q}$ be the polyhedron 
defined by $\Qr:=(P-\what{\vr})\cap \Mb_{\Q}$. 
Then  we have
\begin{equation}
X(M,P)\GIT T^{'}(\vr)
=X\left(\Mb, \Qr \right).
\label{GIT=r}
\end{equation}}
\end{flushleft}

{\it Proof.}

We see that (\ref{GIT=r}) holds when
$\vr\in M^{'}$ and $\what{\vr}\in M$ because upon the shift by 
$\what{\vr}$, each element of $C(P)\cap \tM$ turns to
one of $C(P-\what{\vr})\cap \tM$.

Let $e$ be  the least positive integer 
such that $e\vr\in M^{'}$.
Without loss of generality, we can restrict 
$\what{\vr}\in \pi_{\Q}^{-1}(\vr)$ 
to those which satisfy $e\what{\vr}\in M$.
To deal with this case, 
we use the dilatation invariance of the toric data~:

For a graded ring $G:=\bigoplus_{k\geq 0}G_k$,
define its $e$~th Segre transform $G^{(e)}$ for $e\in \N$ by
$G^{(e)}_k=G_{ek}$ and
$G^{(e)}:=\bigoplus_{k\geq 0}G^{(e)}_k
=\bigoplus_{k\geq 0}G_{ek}$.
Then we have 
\begin{equation}
\Proj G \cong \Proj G^{(e)}. 
\end{equation}
We easily see that the $e$~th Segre transform of $C(P)\cap \tM$
coincides with $C(eP)\cap \tM$, so that
\begin{equation}
X(M,P) \cong X(M,eP).
\label{scaling}
\end{equation}

 Then we have
\begin{align}
X\left( \Mb,(P-\what{\vr})\cap \Mb_{\Q}\right)
&\cong X\left( \Mb,(eP-e\what{\vr})
\cap \Mb_{\Q}\right)
\nonumber \\
&\cong \Proj  
\left(     
C(eP)\cap M
\right)^{T^{'}(e\pmb{r})}.
\label{tempo}
\end{align}

To finish the proof of the Proposition, we have only to 
prove the following lemma~:

\begin{flushleft}
{\it Lemma 2.2.1.}
{\sl The graded ring 
$\left(C(eP)\cap M\right)^{T^{'}(e\pmb{r})}$ 
coincides with 
$\left(C(P)\cap M\right)^{T^{'}(\pmb{r})}$.} 
\end{flushleft}
{\it Proof of Lemma 2.2.1.}

For simplicity, we set temporally 
$G_k:=(C(P)\cap \tM)_k=kP\cap M$,
$G:=C(P)\cap \tM$, and $G^{(e)}:=C(eP)\cap \tM$.
Any element of $G^{T^{'}(\pmb{r})}$ can be written as
$\sum_{j=1}^L (k_j,\vm_j)$,
where the total $T^{'}(\vr)$ charge is
$\sum_{j=1}^L(\pi(\vm_j)-k_j\vr)=\bold{0}$, that is,
$(\sum_{j=1}^L k_j)\skima
\vr=\sum_{j=1}^L \pi(\vm_j)\in M^{'}$,
which implies that 
$\sum_{j=1}^N k_j$,
which is the degree of $\sum_{j=1}^L(k_j,\vm_j)$,
should be a multiple of $e$.
Thus we see that 
if we define a subring $H$ of $G$ by
\begin{align}
H_k&=G_k,\qquad \text{if}\ k\equiv 0 \mod{e}, \nonumber \\
H_k&=0,\qquad \ \ \text{otherwise},\nonumber
\end{align}
then we have $G^{T^{'}(\pmb{r})}=H^{T^{'}(\pmb{r})}$.

Now take an arbitrary element $(ek,\vm)\in H_{ek}=G^{(e)}_k$.
When regarded as an element of $H_{ek}$,
its $T^{'}(\vr)$ charge is $(\pi(\vm)-ek\vr)$, 
which is the same as its  $T^{'}(e\vr)$ charge 
regarded as an element of $G^{(e)}_k$. \hfill $\Box$

Then the combination of (\ref{tempo}) and Lemma~2.2.1
proves the Proposition~2.2.\hfill $\Box$

\subsection{\Ka Moduli Space}
We consider here the $\vr$-dependence 
of the topology, or the {\it phase} in physics terminology,
 of the quotient toric variety (\ref{GIT=r}).
The quotient variety is the toric variety associated with
the normal fan of the polyhedron $\Qr$,
which is given by the {\it slice} 
$P\cap \pi_{\Q}^{-1}(\vr)$ of $P$ translated by $-\what{\vr}$.
Therefore the topology of the quotient variety is determined 
virtually by the shape of the slice $P\cap \pi_{\Q}^{-1}(\vr)$,
which depends on the faces of $P$ that intersect with
the affine subspace $\piQ^{-1}(\vr)$ of $M_{\Q}$.

This observation leads us to define the following 
decomposition of the polyhedron $\piQ(P)$
induced by the $\piQ$-images 
of the faces of $P$ \cite{KSZ}. 
First for each $\vr\in \piQ(P)$,
let $L(\vr)$ be the subset of $\overline{L}(P)$, 
the proper faces of $P$, by
$
L(\vr):=
\left\{
F\in \overline{L}(P)
\left|\
\vr\in \piQ(F)
\right.
\right\}.
$
Then define an equivalence relation $\sim$ 
in $\piQ(P)$ by
$\vr_1\sim \vr_2$ if and only if  
$L(\vr_1)=L(\vr_2)$, 
for $\vr_1,\vr_2\in \piQ(P)$.
We call an equivalence class $K^{0}$ in 
$\piQ(P){/\negthickspace\sim}$
a chamber. 
The polyhedron $\piQ(P)$ 
admits the decomposition
into the disjoint sum of these chambers~:
\begin{equation}
\piQ(P)=\coprod_{K^{0}\in \piQ(P)/\sim} K^{0},
\label{phase}
\end{equation} 
and the topology of the quotient variety $X(\Mb,\Qr)$ 
is constant in each chamber \cite{KSZ}.
Therefore we see that
the decomposition (\ref{phase}) of the parameters space 
$\piQ(P)$ 
represents the phase structure of the toric quotient.

We also define a closed polyhedron $K$ 
to be the closure in $M^{'}_{\Q}$ of the chamber 
$K^0\in \piQ(P){/\negthickspace\sim}$.
Conversely $K^0$ is recovered as
the {\it relative interior} of $K$.

Then the collection of the polyhedra $\kahler$
defined by
\begin{equation}  
\kahler:=\{K|\ K^0\in \piQ(P){/\negthickspace\sim} \}
\label{polyhedral-complex}
\end{equation} 
constitutes 
a {\it polyhedral complex}~\cite[Lecture 5.1]{ziegler}
in $M^{'}_{\Q}$,
which means that for each $K\in \kahler$, 
every face of $K$ belongs to $\kahler$ 
and the intersection $K_1\cap K_2$ 
of any two elements of $\kahler$ 
is the face of both $K_1$ and $K_2$~; in particular
$\kahler$ is a fan if it consists of polyhedral cones,
which is true if $P$ itself is a cone.
We call the polyhedral decomposition of $\piQ(P)$ 
defined by the complex (\ref{polyhedral-complex})
the \Ka moduli space 
associated with the toric quotient.

We define the \Ka walls to be 
the $\piQ$-image of the skeleton of $P$ 
consisting of all the faces of codimensions 
$q+1$.
The \Ka walls is the region 
where the toric quotient construction
{\it degenerates} in the sense that
for each $\vr$ in the \Ka walls,
there is a face $F$ of $P$ such that
$F$ and $\piQ^{-1}(\vr)$ intersect 
despite of the fact that
the sum of their codimensions in $M_{\Q}$
exceeds $p=\dim M_{\Q}$.

We are thus mainly interested 
in the \Ka moduli parameters
in the complement of the \Ka walls in $\piQ(P)$,
which is the disjoint union of the chambers
of the maximal dimensions \cite{thaddeus},
which we call the maximal chambers.
We also call the closure of a maximal chamber
in $M^{'}_{\Q}$ a ``maximal polyhedron''.

 
The $\piQ$-image
of each  face of $P$ of codimensions less than
$q+1$ is a union of several maximal polyhedra.
Let $L(P)^{(k)}$ be the subset of $L(P)$
consisting of the faces of codimensions $k$. 
For each $F\in L(P)^{(k)}$, 
where $k\leq q$,
we can define a $k$-cone in $\Nb_{\Q}$ by
\begin{equation}
\overline{C}(F):=i^*_{\Q}(C(F))
=\pos  
\left\{\skima
\vvb_a
\left|\
a\in I(F)
\right.
\right\},
\end{equation}
where $i^*$ is the lattice surjection
{}from $N$ to $\Nb$ (\ref{exact-N}).

Then we see that for any 
$\vr\in \text{int}\ \piQ(F)$,
the normal fan ${\cal N}(\Qr)$ of the quotient 
has the $k$-cone $\overline{C}(F)$ defined above.
This is because if $\vr\in \text{int}\ \piQ(F)$,
then the slice $P\cap \piQ^{-1}(\vr)$ has the 
face $F\cap \piQ^{-1}(\vr)
=\bigcap_{a\in I(F)}\athfacet
\cap \piQ^{-1}(\vr)$ 
of codimensions $k$,
the normal cone of which is precisely $\overline{C}(F)$.

The following two cases are of particular importance~:
first for the $a$~th facet 
$\athfacet$ and 
for any $\vr\in \piQ(\athfacet)$,
the slice has the facet 
$\athfacet \cap \piQ^{-1}(\vr)$,
so that the normal fan ${\cal N}(\Qr)$ 
has the 1-cone $\pos \{\skima\vvb_a\}$,
which means that the quotient variety has 
the exceptional divisor corresponding to 
$\vvb_a$. 
%
We say two vectors 
$\vvb_a$ and $\vvb_b$
in $\Nb$ to be {\it incompatible} 
if
$\text{int}\ \piQ(\athfacet)$ and 
$\text{int}\ \piQ(\bthfacet)$
have no common point~;
then the two vectors cannot appear 
simultaneously in the quotient fan 
outside the \Ka walls~;
second for $F\in L(P)^{(q)}$ and 
$\vr \in \text{int}\ \piQ(F)$,
the slice has the vertex
$\bigcap_{a\in I(F)}\athfacet
\cap \piQ^{-1}(\vr)$, 
which corresponds to 
the maximal cone 
$\overline{C}(F)\subset \Nb_{\Q}$
of the normal fan.

Because the normal fan ${\cal N}(\Qr)$ 
is determined by listing
its maximal cones,
we obtain the following description
of the phase structure of the quotient variety
outside the \Ka walls.

Let us 
call a subset $S$ of $L(P)^{(q)}$ 
{\it coherent} if the collection 
of the cones in $\Nb_{\Q}$, 
\begin{equation}
\Sigma(S):=
\left\{
L\left(\overline{C}(F)\right) 
\left|\
 F\in S\right.
\right\},
\end{equation} 
defines a fan,
where $L(\overline{C}(F))$, 
the face lattice of $\overline{C}(F)$,
is the set of all the faces of $\overline{C}(F)$,
and if the subspace of $\piQ(P)$ defined by
\begin{equation}
K(S):=\bigcap_{F\in S}\piQ(F)=
\bigcap_{F\in S}
\piQ\left( \cap_{a\in I(F)} 
\athfacet \right)
\label{result}
\end{equation}
has an interior point, that is,
if $K(S)$ is a maximal polyhedron.

\begin{flushleft}
{\it Proposition 2.3.}
{\sl The \Ka moduli space $\kahler$ associated with
the toric quotient is  
\begin{equation}
\kahler=\left\{
L\left(K(S)\right)
\left|\
S\subset L(P)^{(q)}:\  \mathrm{coherent}
\right.
\right\}.
\label{result-1}
\end{equation}}
\end{flushleft}
\begin{flushleft}
{\it Proposition 2.4.}
{\sl For each coherent subset $S\subset L(P)^{(q)}$,
we have
\begin{equation}
X(\Mb,\Qr)\cong X^*(\Nb,\Sigma(S)),
\quad \forall \vr\in \mathrm{int}\ K(S),
\label{result-2}
\end{equation}
where $X^*(\Nb,\Sigma(S))$ is the toric variety 
defined by the fan $\Sigma(S)$.}
\end{flushleft}

Note that (\ref{result-1}) and (\ref{result}) 
generalize the descriptions
of the GKZ secondary fan \cite{GKZ} and its maximal cones
given in \cite[(4.2)]{BFS},
where $M=\Z^p$,
$P=\pos \{\skima\ve_1,\dots, \ve_p\}
\cong ({\Q}_{\geq 0})^p$ is the basic simplicial cone,
and $X(M,P)\cong \C^{p}$,
which has been used in the investigation 
of the \Ka moduli space of {\it bulk string}
compactified on a Calabi--Yau manifold \cite{HLY}.

\single
\section{D-Brane Configuration Space}

\subsection{Calabi--Yau Orbifolds}
Let $\{a_1,\dots,a_d\}$ be a $d$-tuple of the integers,
and $\omega$ be a primitive $n$~th root of unity.
We define $\Ga$ to be a group
isomorphic to the cyclic group $\Z_n:=\Z/n\Z$ 
and define the action of the generator 
$g\in \Ga$ on $\C^d$ by
\begin{equation}
g\cdot x_{\mu}=\omega^{\amu}x_{\mu},
\quad 1\leq \mu \leq d.
\end{equation} 
We denote the quotient space by $\CY$.
The followings are well-known~:
\begin{itemize}
\item
$\CY$ has an isolated singularity at the origin 
if and only if \ $(\amu,n)=1,\quad  \forall \mu$.
\item
$\CY$ is a Calabi--Yau variety if and only if \ 
$\sum_{\mu=1}^d \amu\equiv 0\mod{n}$.
\end{itemize}
We restrict ourselves to the models in which
the orbifold $\CY$ is a Calabi--Yau variety  
with an isolated singularity unless otherwise stated, 
because our main interest is 
the study of the \config\ space ${\cal M}$
of a D-brane localized 
at the singular point of the Calabi--Yau variety $\CY$.
We denote the model characterized by the integers
$(a_1,\dots,a_d;n)$ above by $1/n(a_1,\dots,a_d)$
for simplicity.

Here we give some facts about the Calabi--Yau orbifolds.
An advanced introduction to this subject 
can be found in \cite{DH}.
First of all, $\CY$ is a toric variety.
A useful choice of the dual pair of the lattices 
to describe $\CY$ is the following~:
\begin{align}
\Nb_0 &:=\Z^d+\frac{1}{n}(a_1,\dots,a_d)\ \Z,
\label{N-lattice}\\
\Mb_0 &:=
\left\{
\vm\in \Z^d
\left|\
\vm \cdot \va \equiv 0 \mod{n}
\right.
\right\},\quad \va:=(a_1,\dots,a_d).
\label{M-lattice}
\end{align}
Let $\{\ve^*_1,\dots,\ve_d^*\}$ and
$\{\ve_1,\dots,\ve_d\}$ be the set of 
the fundamental vectors 
of $(\Nb_0)_{\Q}$ and $(\Mb_0)_{\Q}$ respectively, 
which generate the dual pair of simplicial cones~:
\begin{align}
C_0^*&=\pos \{\ve^*_1,\dots,\ve_d^*\}\cong
(\Q_{\geq 0})^d
\subset (\Nb_0)_{\Q}, 
\label{N-cone}\\
C_0&=\pos \{
\ve_1,\dots,\ve_d\}\cong
(\Q_{\geq 0})^d
\subset (\Mb_0)_{\Q}.
\label{M-cone}
\end{align} 
Then we have
\begin{equation}
\CY
=X(\Mb_0,C_0)
=\Spec\skima (\Mb_0\cap C_0)
=X^*(\Nb_0,C^*_0).
\label{affine-orbifold}
\end{equation}
To see this, it suffices to note that
the affine coordinate ring of $\CY$ is
the $\Ga$-invariant part of $\C[x_1,\dots,x_d]$,
which is precisely $(\Mb_0\cap C_0)$.
The simplicial cone $C^*_0\subset (\Nb_0)_{\Q}$
is the fan associated with $\CY$.
Thus a toric blow-up of $\CY$ 
corresponds to a subdivision of the cone $C^*_0$ 
by incorporating new 1-cones,
the primitive vectors of which 
correspond to {\it exceptional divisors}.
For simplicity, we will confuse the primitive vector
of a 1-cone with the exceptional divisor 
associated with it.
Let $T=\conv \{\ve^*_1,\dots,\ve_d^*\}$ 
be the fundamental simplex 
in $(\Nb_0)_{\Q}$ associated with the orbifold.
A primitive vector $\vvb\in \Nb_0$ is classified 
by its age,
which is defined to be the positive integer $k$ 
such that $\vvb\in kT$.
Incorporation of $\vvb\in \Nb_0$ 
in subdivision of the fan $C_0^*$ 
preserves the Calabi--Yau property  
if and only if its age is 1.
Thus a primitive vector of age 1 is said to be
{\it crepant}.
A crepant toric blow-up of $\CY$ corresponds 
to a subdivision of $T$ using lattice points in $T$.
We define the weight vector $\vwb$ 
associated with a primitive vector $\vvb\in \Nb$
by $\vwb:=n\vvb\in \Z^d$.

We can read the physical Hodge numbers 
of bulk string $(h^{p,p})$
``compactified'' on $\CY$ from the Ehrhart series 
for $(\Nb_0,T)$ \cite{BD} as
\begin{equation}
\sum_{k\geq 0}l\left(kT\right)\skima y^k
=\frac{1}{(1-y)^d} \sum_{p=0}^{d-1} h^{p,p}\skima y^p,
\end{equation}
where $l\left(kT\right)$ is the number of the lattice points 
in the dilated simplex $kT$, that is,
$l\left(kT\right)=\text{card}\skima (kT\cap \Nb_0)$.
In particular, the number of the crepant divisors
$h^{1,1}=l(T)-d$ equals to the dimensions 
of the \Ka moduli space of bulk string 
``compacitified'' on $\CY$.

There is a striking difference between 
$d=4$ orbifolds from $d=2,3$ ones~: 
in general, incorporation of the crepant divisors only 
is not enough to resolve $\CY$ completely
into a smooth variety for $d=4$ as opposed to $d=2,3$ cases.

\renewcommand{\theenumi}{\Alph{enumi}}
\renewcommand{\labelenumi}{(\theenumi)}
In \cite{mohri}, we have divided the $d=4$ models 
into the following three classes~:
\begin{enumerate}
\item 
the models that admit a crepant resolution.
\item 
those that have no crepant divisors, the singularities of 
which are called {\it terminal},
consisting of the models of the form~:
$1/n(1,a,n-1,n-1)$
where $(n,a)=1$ \cite{MS}.
\item
those that have at least one crepant divisor, 
but do not admit any crepant resolutions.
\end{enumerate}
The complete identification of the (A) class,
that is, the classification of the isolated 
cyclic quotient {\it Gorenstein} singularities
in four or higher dimensions 
for which crepant resolutions are possible 
is very interesting but unsolved 
mathematical problem \cite{reid}, 
the physical meaning of which is yet to be elucidated.
It is clear that the examples of the (A) class 
shown in \cite{mohri}, 
\begin{equation}
1/(3m+1)(1,1,1,3m-2),\quad
1/(4m)(1,1,2m-1,2m-1), \qquad m\in \N,
\label{trivial}
\end{equation}
are only the tip of the iceberg.
Recently, however, a considerable progress 
in this subject has been made in \cite{DHH,DH}. 
A remarkable new series in the (A) class,
the $m$~th member of which is called the 
{\it 4}-dimensional
{\it geometric progress singularity-series} of ratio $m$
$\left(\text{GPSS}(4;m)\right)$, 
is given in \cite[Conjecture 10.2]{DH}~: 
\begin{flushleft}
{\it Conjecture (Dais--Henk).}
{\sl $1/\{(1+m)(1+m^2)\}(1,m,m^2,m^3)$ model 
admits {\it a} crepant resolution for each $m\in \N$.}
\end{flushleft}
The {\it same} conjecture was also made by the author,
who have only checked that 
the {\it Delaunay triangulation} 
\cite{Del}, \cite[p.~146]{ziegler}, 
of $T$ by the lattice points in it
yields a crepant resolution, 
which is {\it not} unique for $m\geq 3$, up to $m=10$.
%
\subsection{D-Brane Configuration Space}
We consider the configuration space 
of a D1-brane localized at the singular point of $\CY$.
This can be realized as follows~:
First we consider $n=|\Ga|$ D1-branes 
localized at the origin of $\C^d$.
We assign the Chan--Paton indices $i\mod{n}$ to the D1-branes.
Then the world sheet theory on the D1-branes is 
$\text{U}(n)$ gauge theory with $(8,8)$ \susy.
The configuration of the D1-branes is described by the $d$-tuple
of the matrices  $\{ (X_{\mu})^i_j \}$ taking values 
in the adjoint representation of $\text{U}(n)$ \cite{witten2}~;
Second taking into account the $\Ga$-actions 
on the Lorentz indices $(\mu)$ and the Chan--Paton indices $(i)$,
on which $\Ga$ acts as cyclic permutations,
we define the configuration of a D1-brane on the orbifold $\CY$
to be that of $n$ D1-branes on $\C^d$ invariant under  
the simultaneous action of $\Ga$ on the Lorentz 
and the Chan--Paton indices \cite{DGM,DM}.
In the next section, we use a closely related idea 
in the definition of Hilbert schemes of $n$ points on $\C^d$. 

The world sheet \susy\  is reduced, at this point, 
to $(4,4)$, $(2,2)$ and $(0,2)$ for $d=2,3$ and for $d=4$ 
respectively, with the exception that
the \susy\  of $d=4$ (B) model is enhanced to $(0,4)$ 
\cite{mohri}.
We can also consider a model with 
$\sum_{\mu}a_{\mu}\not\equiv 0 \mod{n}$,
where the \susy\  is completely broken \cite{DGM}.
Let $R_a$ be the one dimensional representation of $\Ga$ 
over $\C$ on which the generator $g\in \Ga$ acts 
as multiplication by $\omega^a$.  
Then the D-brane matrices $(X_{\mu})$ take values in
$\left(Q\otimes \text{End}(R)\right)^{\Ga}
\cong \Hom_{\Ga}(R,R\otimes Q)$ 
\cite{kronheimer,sardo,infirri}, 
where the two $\Ga$-modules,
\begin{equation}
R=\bigoplus_{i=1}^{n} R_i,\qquad
Q=\bigoplus_{\mu=1}^d R_{\amu},   \label{regular-rep}
\end{equation}
carry the Chan--Paton and the Lorentz $\Ga$-quantum numbers
of the matrices respectively. 
Note that we have done the discrete Fourier transformation
on the Chan--Paton indices, so that the $\Ga$-action on those
is diagonalized.

To be explicit, the matrix elements that can be nonzero are
\begin{equation}
x_{\mu}^{(i)}:=(X_{\mu})^i_{i+\amu}, 
\label{as-before}
\end{equation}
and the \config\ space of the D1-brane on $\CY$
is the solution space of the following equations~:
\begin{alignat}{2}
&\left[X_{\mu},X_{\nu}\right]=O, &\quad
&\text{F-flatness equation},
\label{M-F-flat}\\
\sum_{\mu=1}^d
&\left[X_{\mu},X_{\mu}^{\dagger}\right]
-\text{diag}(r_1,\dots,r_n)=O, &\quad
&\text{D-flatness equation},
\label{M-D-flat}
\end{alignat}
divided by the action of 
$\text{U}(1)^n/\text{U}(1)_{\text{diag}}$,
where $x_{\mu}^{(i)}$ has the $i$~th $\text{U}(1)$ charge 1
and the $(i+\amu)$~th $\text{U}(1)$ charge $-1$, 
and the others 0 as seen from (\ref{M-D-flat}).

To have a solution to (\ref{M-D-flat}),
the Fayet--Iliopoulos (or \Ka) moduli parameters $\vr:=(r_i)$ 
must satisfy $\sum_{i=1}^n r_i=0$.
 
The F-flatness equation (\ref{M-F-flat})
can be solved as follows \cite{DGM}~:
We redefine the generator of $\Ga$ so that $a_{d}=-1 \mod{n}$.
Then the matrix elements (\ref{as-before}) can be represented 
by $x_d^{(i)}$, $i=1,\dots,n$ and 
$x_{\mu}^{(0)}$, $\mu=1,\dots,d-1$ as
\begin{equation}
x_{\mu}^{(i)}=x_{\mu}^{(0)}\ 
\frac{\prod_{j=1}^{i}x_d^{(j)}
\cdot\prod_{j=1}^{\amu}x_d^{(j)}}%
{\prod_{j=1}^{i+\amu}x_d^{(j)}}.
\label{solution-to-F}
\end{equation}
We see that the solution space 
of the F-flatness equation (\ref{M-F-flat}), 
which we denote by $\A$, is the $(n-1+d)$-dimensional
affine variety embedded in $\C^{nd}$ 
defined by the equations of monomial type
(\ref{solution-to-F}), 
which shows that $\A$ is a toric variety.
The \config\ space of the D1-brane, which we denote by $\Mod$,
is also toric because it is obtained as a toric quotient 
of $\A$ (\ref{M-D-flat}).

In the next subsection, we give a toric description of $\Mod$,
based on the formalism developed in the last section,
which elucidates the structure of the \Ka moduli space 
associated with the toric quotient 
$\A{\GIT}(\C^*)^{n-1}(\vr)$, 
as well as provides us with an efficient method
to compute the \config\ space $\Mod$
for any $\vr\in \Q^{n-1}$.

\subsection{Toric Description of the D-Brane Configuration Space}
According to (\ref{solution-to-F}), we propose the following
toric description of $\A$ \cite{infirri}~:
Let $\Mtot$ be a lattice of rank $nd$
generated by $\ve_{\mu}^{(i)}$, 
$0\leq i \leq n-1$, $1\leq \mu\leq d$,
and $\Msub$ be the sublattice of rank $(n-1)(d-1)$
of $\Mtot$ generated by
\begin{equation}
\vf^{(i)}_{\mu}:=\ve_{\mu}^{(i)}-\ve_{\mu}^{(0)}
+\sum_{j=1}^{i+\amu}\ve_{d}^{(j)}
-\sum_{j=1}^{i}\ve_{d}^{(j)}
-\sum_{j=1}^{\amu}\ve_{d}^{(j)},
\quad  \mu\ne d,\ i\ne 0,
\end{equation}
with the injection $j:\Msub\ra \Mtot$.
Let $M=\Mtot/\Msub$
be the quotient lattice  
of rank $n-1+d$ and
$p:\Mtot\ra M$ be the projection.
If we define $\cone$ to be the basic simplicial cone 
in $\Mtot_{\Q}$, that is,
\begin{equation}
\left.
\cone=\pos 
\left\{\skima
\ve_{\mu}^{(i)}
\right|\ 
1\leq \mu\leq d,\ 
0\leq i\leq n-1
\right\},
\end{equation}
then its $p_{\Q}$-image
$P:=p_{\Q}(\cone)$ is a cone in $M_{\Q}$ 
and we have~\cite{infirri}
\begin{equation}
\A=X(M,P)=\Spec \left(P\cap M\right).
\end{equation}
The D-brane \config\ space $\Mod$ 
can be realized as the toric quotient 
of $\A$ as follows \cite{infirri}~:
Let $M^{'}\subset \Z^n$ be 
the lattice of rank $n-1$ 
generated by $\ve_i-\ve_{i+1}$, 
$1\leq i \leq  n-1$,
where $(\ve_i)$ is the generators of $\Z^n$,
and $\pi^{'}:\Mtot\ra M^{'}$ the lattice projection
\begin{equation}
\pi^{'}(\ve_{\mu}^{(i)}):=\ve_i-\ve_{i+\amu},
\end{equation}
which is determined according to the $\text{U}(1)^n$
charge assignment of $x_{\mu}^{(i)}$.
It is easily seen that $\pi^{'}$ factors through
$p$, that is,
there is a projection 
$\pi:M\ra M^{'}$ such that
$\pi^{'}=\pi\circ p$. 
Finally we define a sublattice of rank $d$ 
of $M$ by
$\Mb:= \Ker \pi
\cong \Ker  \pi^{'}/\text{Im}\ j$. 
Note that $\pi_{\Q}(P)=M^{'}_{\Q}$.
Then we have
\begin{equation}
\Mod:=X(M,P)\GIT T^{'}(\vr)=X(\Mb,\Qr),
\end{equation}
where $T^{'}$ is the subtorus of $T$ associated with
the sublattice $N^{'}=(M^{'})^*\subset N=M^*$, 
and we regard the Fayet--Iliopoulos parameter 
$\vr$ as a point of $M^{'}_{\Q}$.
We can obtain the fan of $\Mod$ as the normal fan of
the $d$-polyhedron $\Qr$.

\begin{flushleft}
{\it Remark.} 
{\sl It may be confusing to have two lattices of rank $d$,
both of which are associated with the \config\ space $\Mod$~:
$\Mb$ symbolically represents the lattice for a general
quotient toric variety (\ref{exact-M})~;
on the other hand, $\Mb_0$, 
which was originally introduced 
as a useful lattice to describe the orbifold 
$\CY$ in (\ref{M-lattice}), 
is also suited for its blow-up $\Mod$.
Our intention is that  
we use $\Mb_0$ for the concrete descriptions 
of the toric data of $\Mod$ below.}     
\end{flushleft}
\vspace{1cm}

\setlength{\unitlength}{1mm}
\begin{picture}(115,95)(0,15)
\put(81,43){$\pi$}        
\put(85,55){$\pi^{'}$}       
\put(66,51){$p$}       
\put(67,73){$j$}           
\put(59,43){$i$}             
\put(26,40){0}
\put(31,41){\vector(1,0){15}}
\put(48,40){$\Mb$}            
\put(53,41){\vector(1,0){15}}
\put(70,40){$M$}                       
\put(75,41){\vector(1,0){15}}
\put(92,40){$M^{'}$}                  
\put(97,41){\vector(1,0){15}}
\put(114,40){0}
\put(72,38){\vector(0,-1){15}}
\put(71,18){0}
\put(70,62){$\Mtot$}                     
\put(72,60){\vector(0,-1){15}}
\put(70,84){$\Msub$}
\put(72,82){\vector(0,-1){15}}
\put(72,105){\vector(0,-1){15}}
\put(71,107){0}
\put(77,60){\vector(1,-1){15}}
\end{picture}
%
\begin{center}
{\bf Figure 1.} \ Sequences of Lattices
\end{center}
%

\subsection{Some Properties of the \Ka Moduli Space}
We define the action of a generator of $\Ga$ 
on the Chan--Paton indices by
$\g(i):=i+1\pmod{n}$, 
which can be extended to an action of $\Ga$ 
as an automorphism on each lattice shown in Figure 1
in such a manner that any lattice 
homomorphism in Figure 1 becomes $\Ga$-equivariant,
which we denote by
$\g^{(1)},\g^{(0)},\g,\g^{'},\overline{\g}$
for 
$\Msub,\Mtot,M,M^{'},\Mb$ 
respectively. 
For example, the action on the generators of $\Mtot$ 
reads as follows~:
\begin{equation}
\g^{(0)}(\ve_{\mu}^{(i)})=\ve_{\mu}^{(i+1)},
\label{Ga-action}
\end{equation}
while the action on those of $\Msub$ is
\begin{equation}
\g^{(1)}(\vf_{\mu}^{(i)})
=\vf_{\mu}^{(i+1)}-\vf_{\mu}^{(1)}.
\end{equation}
The \Ka moduli space of $1/n(a_1,\dots,a_d)$,
which we denote by
$\kahler^n(a_1,\dots,a_d)$,
is the complete fan in $M^{'}_{\Q}$
obtained as the subdivision of $M^{'}_{\Q}$ induced by 
the $\pi_{\Q}$-images of the faces 
of the cone $P$ in $M_{\Q}$.

The Propositions 3.1--3 stated below 
are immediate consequences of our definitions~:
\begin{flushleft}
{\it Proposition 3.1.}
{\sl If $\{a_1,\dots, a_d\}
=\{b_1,\dots, b_e \}$
as {\it sets}, 
then the two models
$1/n(a_1,\dots,a_d)$ and 
$1/n(b_1,\dots, b_e)$
have the {\it same} \Ka moduli space, that is,
\begin{equation}
\kahler^n(a_1,\dots,a_d)
=\kahler^n(b_1,\dots, b_e),
\label{reduction}
\end{equation}
where the two models above need not 
necessarily satisfy the Calabi--Yau condition.}
\end{flushleft}

We say that the $d$-fold model
$1/n(a_1,\dots,a_d)$ can be reduced to $e$~dimensions,
when (\ref{reduction}) occurs with $d>e$.

\begin{flushleft}
{\it Example.}
We have the reductions of the Calabi--Yau four-fold models
to two~dimensions according to the following identifications~:
\begin{align}
\kahler^m(1,1,m-1,m-1)&=
\kahler^m(1,m-1),     \label{4-1}      \\
\kahler^{3m+1}(1,1,1,3m-2)&=
\kahler^{3m+1}(1,3m-2), \label{4-2}         \\
\kahler^{4m}(1,1,2m-1,2m-1)&=
\kahler^{4m}(1,2m-1). \label{4-3}
\end{align}
\end{flushleft}

A toric two-fold has the virtue
that the listing of the 1-cones alone 
determines its fan.
The four-fold models entering in (\ref{4-1}--\ref{4-3}) 
inherit this property
{}from the corresponding two-fold models,
which considerably simplifies 
the analysis of the \Ka moduli space 
of these four-fold models. 

Let us take $1/m(1,m-1)$ model.
The maximal chambers of the \Ka moduli space 
$\kahler^m(1,m-1)$ can identified 
with the Weyl chambers of $\text{SU}(m)$ \cite{kronheimer},
in which  the D-brane \config\ space
$\Mod$ is in the minimal blow-up phase.  
Correspondingly, the phase of the four-fold
$1/m(1,1,m-1,m-1)$ 
in the Weyl chambers turns out to be 
the non-Calabi--Yau smooth phase 
with Euler number $4(m-1)$,
the fan of which is given 
by the following collection of the maximal cones~:
\begin{align}
&\langle 1,3,4,5 \rangle, \ 
\langle 2,3, 4,5 \rangle, \  
\langle 1,2, 3,m+3 \rangle, \  
\langle 1,2, 4,m+3 \rangle,
 \\ 
&\langle 1,3,l,l+1 \rangle,  \  
\langle 1,4,l,l+1 \rangle,   \ 
\langle 2,3, l,l+1 \rangle,   \  
\langle 2,4,l,l+1 \rangle,
\ \  5\leq l\leq m+2,\nonumber
\end{align}
where the weight vectors above are given by
\begin{align}
&\vwb_1=(m,0,0,0), \ 
\vwb_2=(0,m,0,0), \ 
\vwb_3=(0,0,m,0), \ 
\vwb_4=(0,0,0,m), \nonumber \\
&\vwb_l=(l-4,l-4,m+4-l,m+4-l), \qquad 5\leq l \leq m+3.
\end{align}
Here our convention for the expression  
of the maximal cone \cite{HLY} is~: 
\begin{equation}
\langle s_1,\dots,s_k\rangle:=
\pos \{\skima
\vwb_{s_1},\dots,\vwb_{s_k}\}\subset (\Nb_0)_{\Q}.
\end{equation}
Brute force calculations for $m=2,3$ cases 
can be found in \cite[Section 6.2]{mohri}.

As for $1/(4m)(1,1,2m-1,2m-1)$ model and
$1/(3m+1)(1,1,1,3m-2)$ model,
the D-brane \config\ space $\Mod$
of the four-fold model is 
in the smooth Calabi--Yau phase
if and only if 
that of the corresponding two-fold model is 
in the minimal blow-up phase~;
the former is
in the non-Calabi--Yau smooth phases 
if and only if the latter is
in the non-minimal blow-up phases. 

In the same way, a {\it two-parameter model}~: 
$1/n(1,\dots,1,a,b)$ treated in \cite{DHH}
is one which can be reduced to three dimensions.

\begin{flushleft}
{\it Proposition 3.2.}
{\sl The polyhedron $P$ admits an action of $\Ga$, that is,
$\g_{\Q}(P)=P$.}

{\it Corollary 3.2.1.} 
{\sl The set of the facets of $P$,
which we previously denoted by 
$L(P)^{(1)}=\{ \athfacet|\ a\in \facetset \}$,
are decomposed into $\Ga$-orbits.}
\end{flushleft}
Within each $\Ga$-orbit, the facets share 
a common weight vector for the quotient toric variety.
Each model has the following $d$ $\Ga$-singlets
\begin{equation}
{\cal F}_{\mu}:=
\pos
\left.
\left\{
\skima \vp_{\nu}^{(i)}
\right|\ 
\nu\ne \mu
\right\},
\quad 1\leq \mu \leq d,
\end{equation}
where we set $p(\ve_{\mu}^{(i)})=\vp_{\mu}^{(i)}\in M$ 
for simplicity. 
The weight vector associated with ${\cal F}_{\mu}$ is
$\vwb_{\mu}=n\ve_{\mu}$ for $1\leq \mu\leq d$.

The remaining $\Ga$-orbits are denoted by
\begin{equation}
\left.
\left\{ 
{\cal F}_k^{(j)}:=\g_{\Q}^j
\left(
{\cal F}_{k}^{(0)}
\right)
\right|\ 
0\leq j \leq m_k-1
\right\},
\quad k\geq d+1,
\label{Ga-orbit}
\end{equation} 
where $m_k$ is the length of the $k$~th $\Ga$-orbit,
and we denote the weight vector associated with
the $k$~th orbit by $\vwb_k$, 
which we call the $k$~th exceptional divisor.
\begin{flushleft}
{\it Example.}
For $1/5(1,2,3,4)$ model, the exceptional divisors are
\begin{alignat}{2}
\vwb_5   &=(1,2,3,4), &\qquad
\vwb_6   &=(2,4,1,3)\nonumber \\
\vwb_7   &=(3,1,4,2), &\qquad
\vwb_8   &=(4,3,2,1)\nonumber \\
\vwb_9   &=(4,3,2,6), &\qquad
\vwb_{10}&=(2,4,6,3)\nonumber \\
\vwb_{11}&=(3,6,4,2), &\qquad
\vwb_{12}&=(6,2,3,4).
\end{alignat} 
To describe $k$~th $\Ga$-orbit, 
it suffices to show its 0~th member
${\cal F}_k^{(0)}$ as in (\ref{Ga-orbit}).
Then we have for the age 2 exceptional divisors
\begin{align}
{\cal F}_{5}^{(0)}&=
\pos
\left\{
\skima
\vp_{1}^{(1)},
\vp_{1}^{(2)},
\vp_{1}^{(3)},
\vp_{1}^{(4)},
\vp_{2}^{(1)},
\vp_{2}^{(2)},
\vp_{2}^{(3)},
\vp_{3}^{(1)},
\vp_{3}^{(2)},
\vp_{4}^{(1)}
\right\},\\
{\cal F}_{6}^{(0)}&=
\pos
\left\{
\skima
\vp_{1}^{(1)},
\vp_{1}^{(3)},
\vp_{1}^{(4)},
\vp_{2}^{(3)},
\vp_{3}^{(1)},
\vp_{3}^{(2)},
\vp_{3}^{(3)},
\vp_{3}^{(4)},
\vp_{4}^{(1)},
\vp_{4}^{(3)}
\right\},\\
{\cal F}_{7}^{(0)}&=
\pos
\left\{
\skima
\vp_{1}^{(0)},
\vp_{1}^{(2)},
\vp_{2}^{(0)},
\vp_{2}^{(1)},
\vp_{2}^{(2)},
\vp_{2}^{(4)},
\vp_{3}^{(0)},
\vp_{4}^{(0)},
\vp_{4}^{(2)},
\vp_{4}^{(4)}
\right\},\\
{\cal F}_{8}^{(0)}&=
\pos
\left\{
\skima
\vp_{1}^{(0)},
\vp_{2}^{(0)},
\vp_{2}^{(4)},
\vp_{3}^{(0)},
\vp_{3}^{(3)},
\vp_{3}^{(4)},
\vp_{4}^{(0)},
\vp_{4}^{(2)},
\vp_{4}^{(3)},
\vp_{4}^{(4)}
\right\},
\end{align}
and for the age 3 exceptional divisors
\begin{align}
{\cal F}_{9}^{(0)}&=
\pos
\left\{
\skima
\vp_{1}^{(2)},
\vp_{1}^{(4)},
\vp_{2}^{(1)},
\vp_{2}^{(4)},
\vp_{3}^{(0)},
\vp_{3}^{(2)},
\vp_{3}^{(4)},
\vp_{4}^{(4)}
\right\},\\
{\cal F}_{10}^{(0)}&=
\pos
\left\{
\skima
\vp_{1}^{(2)},
\vp_{1}^{(3)},
\vp_{1}^{(4)},
\vp_{2}^{(2)},
\vp_{2}^{(3)},
\vp_{3}^{(2)},
\vp_{4}^{(1)},
\vp_{4}^{(2)}
\right\},\\
{\cal F}_{11}^{(0)}&=
\pos
\left\{
\skima
\vp_{1}^{(0)},
\vp_{1}^{(1)},
\vp_{2}^{(0)},
\vp_{3}^{(0)},
\vp_{3}^{(4)},
\vp_{4}^{(0)},
\vp_{4}^{(3)},
\vp_{4}^{(4)}
\right\},\\
{\cal F}_{12}^{(0)}&=
\pos
\left\{
\skima
\vp_{1}^{(0)},
\vp_{2}^{(0)},
\vp_{2}^{(2)},
\vp_{2}^{(4)},
\vp_{3}^{(0)},
\vp_{3}^{(3)},
\vp_{4}^{(0)},
\vp_{4}^{(2)}
\right\}.
\end{align}
The action of $\Ga$ on a facet is as follows~:
\begin{equation}
\g_{\Q}\left({\cal F}_{5}^{(0)}\right)
=\pos
\left\{
\skima
\vp_{1}^{(2)},
\vp_{1}^{(3)},
\vp_{1}^{(4)},
\vp_{1}^{(0)},
\vp_{2}^{(2)},
\vp_{2}^{(3)},
\vp_{2}^{(4)},
\vp_{3}^{(2)},
\vp_{3}^{(3)},
\vp_{4}^{(2)}
\right\}.
\end{equation}
We see that the length of each $\Ga$-orbits above is {\tt 5}.
\end{flushleft}
\begin{flushleft}
{\it Example.} For the case of $(1/12)(1,1,5,5)$ model,
the exceptional divisors and 
the lengths of the $\Ga$-orbits are as follows~:
\begin{alignat}{2}
&\vwb_5=(1,1,5,5): &\quad &\text{\tt 12}\nonumber \\
&\vwb_6=(3,3,3,3): &\quad &\text{\tt 12}+\text{\tt 12}
               +\text{\tt 12}+\text{\tt 4}\nonumber \\
&\vwb_7=(5,5,1,1): &\quad &\text{\tt 12} \\
&\vwb_8=(4,4,8,8): &\quad 
&\text{\tt 6}+\text{\tt 6}+\text{\tt 3}\nonumber \\
&\vwb_9=(8,8,4,4): &\quad 
&\text{\tt 6}+\text{\tt 6}+\text{\tt 3}.\nonumber
\end{alignat}
The representatives of $\Ga$-orbits 
for the crepant divisors are
\begin{align}
{\cal F}_{5}^{(0)}&=
\pos
\left.
\left\{ \skima
\begin{aligned}
&\vp_{\mu}^{(1)},
\vp_{\mu}^{(2)},
\vp_{\mu}^{(3)},
\vp_{\mu}^{(4)},
\vp_{\mu}^{(5)},
\vp_{\mu}^{(6)},
\vp_{\mu}^{(7)},
\vp_{\mu}^{(8)},
\vp_{\mu}^{(9)},
\vp_{\mu}^{(10)},
\vp_{\mu}^{(11)}\\
&\vp_{\nu}^{(1)},
\vp_{\nu}^{(2)},
\vp_{\nu}^{(4)},
\vp_{\nu}^{(6)},
\vp_{\nu}^{(7)},
\vp_{\nu}^{(9)},
\vp_{\nu}^{(11)}
\end{aligned} 
\right|
\begin{gathered}
 \mu=1,2\\
\nu=3,4
\end{gathered}
\right\}\nonumber \\
{}^1{\cal F}_{6}^{(0)}&=
\pos
\left.
\left\{ \skima
\begin{aligned}
&\vp_{\mu}^{(0)},
\vp_{\mu}^{(1)},
\vp_{\mu}^{(2)},
\vp_{\mu}^{(5)},
\vp_{\mu}^{(6)},
\vp_{\mu}^{(8)},
\vp_{\mu}^{(9)},
\vp_{\mu}^{(10)},
\vp_{\mu}^{(11)}\\
&\vp_{\nu}^{(1)},
\vp_{\nu}^{(2)},
\vp_{\nu}^{(4)},
\vp_{\nu}^{(5)},
\vp_{\nu}^{(6)},
\vp_{\nu}^{(8)},
\vp_{\nu}^{(9)},
\vp_{\nu}^{(10)},
\vp_{\nu}^{(11)}
\end{aligned} 
\right|
\begin{gathered}
 \mu=1,2\\
\nu=3,4
\end{gathered}
\right\}\nonumber \\
{}^2{\cal F}_{6}^{(0)}&=
\pos
\left.
\left\{ \skima
\begin{aligned}
&\vp_{\mu}^{(3)},
\vp_{\mu}^{(4)},
\vp_{\mu}^{(5)},
\vp_{\mu}^{(6)},
\vp_{\mu}^{(7)},
\vp_{\mu}^{(8)},
\vp_{\mu}^{(9)},
\vp_{\mu}^{(10)},
\vp_{\mu}^{(11)}\\
&\vp_{\nu}^{(1)},
\vp_{\nu}^{(2)},
\vp_{\nu}^{(3)},
\vp_{\nu}^{(4)},
\vp_{\nu}^{(6)},
\vp_{\nu}^{(7)},
\vp_{\nu}^{(8)},
\vp_{\nu}^{(9)},
\vp_{\nu}^{(11)}
\end{aligned} 
\right|
\begin{gathered}
 \mu=1,2\\
\nu=3,4
\end{gathered}
\right\}\nonumber  \\
{}^3{\cal F}_{6}^{(0)}&=
\pos
\left.
\left\{ \skima
\begin{aligned}
&\vp_{\mu}^{(0)},
\vp_{\mu}^{(2)},
\vp_{\mu}^{(3)},
\vp_{\mu}^{(6)},
\vp_{\mu}^{(7)},
\vp_{\mu}^{(8)},
\vp_{\mu}^{(9)},
\vp_{\mu}^{(10)},
\vp_{\mu}^{(11)}\\
&\vp_{\nu}^{(0)},
\vp_{\nu}^{(2)},
\vp_{\nu}^{(3)},
\vp_{\nu}^{(5)},
\vp_{\nu}^{(6)},
\vp_{\nu}^{(7)},
\vp_{\nu}^{(8)},
\vp_{\nu}^{(10)},
\vp_{\nu}^{(11)}
\end{aligned} 
\right|
\begin{gathered}
 \mu=1,2\\
\nu=3,4
\end{gathered}
\right\} \\
{}^4{\cal F}_{6}^{(0)}&=
\pos
\left.
\left\{ \skima
\begin{aligned}
&\vp_{\mu}^{(0)},
\vp_{\mu}^{(1)},
\vp_{\mu}^{(2)},
\vp_{\mu}^{(4)},
\vp_{\mu}^{(5)},
\vp_{\mu}^{(6)},
\vp_{\mu}^{(8)},
\vp_{\mu}^{(9)},
\vp_{\mu}^{(10)}\\
&\vp_{\nu}^{(0)},
\vp_{\nu}^{(1)},
\vp_{\nu}^{(2)},
\vp_{\nu}^{(4)},
\vp_{\nu}^{(5)},
\vp_{\nu}^{(6)},
\vp_{\nu}^{(8)},
\vp_{\nu}^{(9)},
\vp_{\nu}^{(10)}
\end{aligned} 
\right|
\begin{gathered}
 \mu=1,2\\
\nu=3,4
\end{gathered}
\right\}\nonumber \\
{\cal F}_{7}^{(0)}&=
\pos
\left.
\left\{ \skima
\begin{aligned}
&\vp_{\mu}^{(5)},
\vp_{\mu}^{(6)},
\vp_{\mu}^{(7)},
\vp_{\mu}^{(8)},
\vp_{\mu}^{(9)},
\vp_{\mu}^{(10)},
\vp_{\mu}^{(11)}\\
&
\vp_{\nu}^{(1)},
\vp_{\nu}^{(2)},
\vp_{\nu}^{(3)},
\vp_{\nu}^{(4)},
\vp_{\nu}^{(5)},
\vp_{\nu}^{(6)},
\vp_{\nu}^{(7)},
\vp_{\nu}^{(8)},
\vp_{\nu}^{(9)},
\vp_{\nu}^{(10)},
\vp_{\nu}^{(11)}
\end{aligned} 
\right|
\begin{gathered}
 \mu=1,2\\
\nu=3,4
\end{gathered}
\right\},\nonumber 
\end{align}
while those for the age 2 divisors are
\begin{align}
{}^1{\cal F}_{8}^{(0)}&=
\pos
\left.
\left\{ \skima
\begin{aligned}
&\vp_{\mu}^{(0)},
\vp_{\mu}^{(1)},
\vp_{\mu}^{(2)},
\vp_{\mu}^{(4)},
\vp_{\mu}^{(6)},
\vp_{\mu}^{(7)},
\vp_{\mu}^{(8)},
\vp_{\mu}^{(10)}\\
&\vp_{\nu}^{(0)},
\vp_{\nu}^{(4)},
\vp_{\nu}^{(6)},
\vp_{\nu}^{(10)}
\end{aligned} 
\right|
\begin{gathered}
 \mu=1,2\\
\nu=3,4
\end{gathered}
\right\}\nonumber \\
{}^2{\cal F}_{8}^{(0)}&=
\pos
\left.
\left\{ \skima
\begin{aligned}
&\vp_{\mu}^{(0)},
\vp_{\mu}^{(3)},
\vp_{\mu}^{(4)},
\vp_{\mu}^{(5)},
\vp_{\mu}^{(6)},
\vp_{\mu}^{(9)},
\vp_{\mu}^{(10)},
\vp_{\mu}^{(11)}\\
&\vp_{\nu}^{(2)},
\vp_{\nu}^{(3)},
\vp_{\nu}^{(8)},
\vp_{\nu}^{(9)}
\end{aligned} 
\right|
\begin{gathered}
 \mu=1,2\\
\nu=3,4
\end{gathered}
\right\}\nonumber \\
{}^3{\cal F}_{8}^{(0)}&=
\pos
\left.
\left\{ \skima
\begin{aligned}
&\vp_{\mu}^{(0)},
\vp_{\mu}^{(1)},
\vp_{\mu}^{(3)},
\vp_{\mu}^{(4)},
\vp_{\mu}^{(6)},
\vp_{\mu}^{(7)},
\vp_{\mu}^{(9)},
\vp_{\mu}^{(10)}\\
&\vp_{\nu}^{(0)},
\vp_{\nu}^{(3)},
\vp_{\nu}^{(6)},
\vp_{\nu}^{(9)}
\end{aligned} 
\right|
\begin{gathered}
 \mu=1,2\\
\nu=3,4
\end{gathered}
\right\} \\
{}^1{\cal F}_{9}^{(0)}&=
\pos
\left.
\left\{ \skima
\begin{aligned}
&\vp_{\mu}^{(3)},
\vp_{\mu}^{(4)},
\vp_{\mu}^{(9)},
\vp_{\mu}^{(10)}\\
&\vp_{\nu}^{(0)},
\vp_{\nu}^{(1)},
\vp_{\nu}^{(2)},
\vp_{\nu}^{(3)},
\vp_{\nu}^{(6)},
\vp_{\nu}^{(7)},
\vp_{\nu}^{(8)},
\vp_{\nu}^{(9)}
\end{aligned} 
\right|
\begin{gathered}
 \mu=1,2\\
\nu=3,4
\end{gathered}
\right\}\nonumber  \\
{}^2{\cal F}_{9}^{(0)}&=
\pos
\left.
\left\{ \skima
\begin{aligned}
&\vp_{\mu}^{(0)},
\vp_{\mu}^{(2)},
\vp_{\mu}^{(6)},
\vp_{\mu}^{(8)}\\
&\vp_{\nu}^{(0)},
\vp_{\nu}^{(2)},
\vp_{\nu}^{(4)},
\vp_{\nu}^{(5)},
\vp_{\nu}^{(6)},
\vp_{\nu}^{(8)},
\vp_{\nu}^{(10)},
\vp_{\nu}^{(11)}
\end{aligned} 
\right|
\begin{gathered}
 \mu=1,2\\
\nu=3,4
\end{gathered}
\right\}\nonumber \\
{}^3{\cal F}_{9}^{(0)}&=
\pos
\left.
\left\{ \skima
\begin{aligned}
&\vp_{\mu}^{(2)},
\vp_{\mu}^{(5)},
\vp_{\mu}^{(8)},
\vp_{\mu}^{(11)}\\
&\vp_{\nu}^{(1)},
\vp_{\nu}^{(2)},
\vp_{\nu}^{(4)},
\vp_{\nu}^{(5)},
\vp_{\nu}^{(7)},
\vp_{\nu}^{(8)},
\vp_{\nu}^{(10)},
\vp_{\nu}^{(11)}
\end{aligned} 
\right|
\begin{gathered}
 \mu=1,2\\
\nu=3,4
\end{gathered}
\right\}. \nonumber
\end{align}
Note that the reducibility of this model
to two dimensions (\ref{4-3})
is reflected in the structure of the facets of $P$.
\end{flushleft}

\begin{flushleft}
{\it Proposition 3.3.}
{\sl $\Ga$ acts on $M^{'}_{\Q}$ 
as an symmetry of the toric quotient~:
\begin{equation}
{\cal M}(\vr)\cong {\cal M}(\g^{'}_{\Q}(\vr)), 
\quad \vr\in M^{'}_{\Q}.
\end{equation}}
\end{flushleft}

\subsection{Phases of Calabi--Yau Four-Fold Models}
Here we describe some typical phases of
Calabi--Yau four-fold  models, 
leaving the cases of $d=3$ for the reader's exercise.

In this subsection, we identify 
$M^{'}
=\{\skima \vr\in \Z^n |\sum_{i=0}^{n-1}\skima r_i=0 \}$ 
with $\Z^{n-1}$ by discarding its zeroth component $r_0$.
\subsubsection{(1/12)(1,1,5,5) model}
First of all, we need to choose the representative
of the facets for each weight vector $\vwb_k$ 
for $k=5,\dots,9$, which we denote by ${\cal F}_k$, 
so that they are compatible, that is, 
\begin{equation}
\bigcap_{k=5}^9 \pi_{\Q}\left({\cal F}_k\right)
\end{equation}
is a $4$ dimensional cone in $M^{'}_{\Q}$. 
Our choice is as follows~:
\begin{equation}
{\cal F}_5={\cal F}_5^{(3)},\quad
{\cal F}_6={}^4{\cal F}_6^{(0)},\quad
{\cal F}_7={\cal F}_7^{(1)},\quad   
{\cal F}_8={}^1{\cal F}_8^{(0)},\quad
{\cal F}_9={}^2{\cal F}_9^{(0)}.
\end{equation}
Consider the following candidates of the phases 
realized in maximal cones of the \Ka moduli space
$\kahler^{12}(1,1,5,5)$~:
\vspace{0.5cm}

phase I ($\Sigma_{\text{I}}$)
\begin{alignat}{4}
&\langle 1,2,3,7 \rangle, &\quad
&\langle 1,2,4,7 \rangle, &\quad
&\langle 1,3,4,5  \rangle, &\quad
&\langle 2,3,4,5  \rangle, \nonumber \\
&\langle 2,4,5,6 \rangle, &\quad
&\langle 2,4,6,7  \rangle, &\quad
&\langle 1,4,6,7  \rangle, &\quad
&\langle 2,3,6,7  \rangle,           \\
&\langle 1,4,5,6  \rangle, &\quad
&\langle 1,3,6,7  \rangle, &\quad
&\langle 2,3,5,6  \rangle, &\quad
&\langle 1,3,5,6  \rangle. \nonumber 
\end{alignat} 

phase II ($\Sigma_{\text{II}}$)
\begin{alignat}{4}
&\langle 1,2,3,7 \rangle, &\quad
&\langle 1,2,4,7 \rangle, &\quad
&\langle 1,3,4,5  \rangle, &\quad
&\langle 2,3,4,5  \rangle, \nonumber \\
&\langle 2,4,6,8 \rangle, &\quad
&\langle 1,4,6,7  \rangle, &\quad
&\langle 2,4,5,8  \rangle, &\quad
&\langle 1,4,5,8  \rangle,      \\
&\langle 1,4,6,8  \rangle, &\quad
&\langle 1,3,5,8  \rangle, &\quad
&\langle 2,3,6,8  \rangle, &\quad
&\langle 1,3,6,8  \rangle, \nonumber \\ 
&\langle 2,4,6,7  \rangle, &\quad
&\langle 2,3,5,8  \rangle, &\quad
&\langle 2,3,6,7  \rangle, &\quad
&\langle 1,3,6,7  \rangle. \nonumber 
\end{alignat} 

phase III ($\Sigma_{\text{III}}$)
\begin{alignat}{4}
&\langle 1,2,3,7 \rangle, &\quad
&\langle 1,2,4,7 \rangle, &\quad
&\langle 1,3,4,5  \rangle, &\quad
&\langle 2,3,4,5  \rangle, \nonumber \\
&\langle 1,4,6,9 \rangle, &\quad
&\langle 2,4,5,6  \rangle, &\quad
&\langle 2,4,6,9  \rangle, &\quad
&\langle 2,4,7,9  \rangle,      \\
&\langle 1,4,7,9  \rangle, &\quad
&\langle 1,3,6,9  \rangle, &\quad
&\langle 1,3,7,9  \rangle, &\quad
&\langle 1,4,5,6  \rangle, \nonumber \\ 
&\langle 2,3,7,9  \rangle, &\quad
&\langle 2,3,6,9  \rangle, &\quad
&\langle 2,3,5,6  \rangle, &\quad
&\langle 1,3,5,6  \rangle. \nonumber 
\end{alignat} 

phase IV ($\Sigma_{\text{IV}}$)
\begin{alignat}{4}
&\langle 1,2,3,7 \rangle, &\quad
&\langle 1,2,4,7 \rangle, &\quad
&\langle 1,3,4,5  \rangle, &\quad
&\langle 2,3,4,5  \rangle, \nonumber \\
&\langle 2,4,6,8 \rangle, &\quad
&\langle 2,4,5,8  \rangle, &\quad
&\langle 2,4,6,9  \rangle, &\quad
&\langle 1,4,6,9  \rangle,  \nonumber \\
&\langle 2,4,7,9  \rangle, &\quad
&\langle 1,4,6,8  \rangle, &\quad
&\langle 1,4,7,9  \rangle, &\quad
&\langle 1,3,6,9  \rangle,           \\
&\langle 2,3,7,9 \rangle, &\quad
&\langle 1,3,7,9  \rangle, &\quad
&\langle 1,4,5,8  \rangle, &\quad
&\langle 2,3,6,9  \rangle, \nonumber \\
&\langle 2,3,6,8  \rangle, &\quad
&\langle 2,3,5,8  \rangle, &\quad
&\langle 1,3,5,8  \rangle, &\quad
&\langle 1,3,6,8   \rangle. \nonumber 
\end{alignat} 
Here $\Sigma_{\text{I--IV}}$ means 
the corresponding fan.
The phase I is the smooth Calabi--Yau phase~;
the phase II--IV are non-Calabi--Yau smooth phases,
which are blow-ups of the phase I.

\setlength{\unitlength}{1mm}
\begin{picture}(80,40)(5,5)
\put(9,19){IV}
\put(29,10){III}
\put(29,28){II}
\put(50,19){I}
\put(14,23){\vector(2,1){13}}
\put(14,19){\vector(2,-1){13}}
\put(35,29){\vector(2,-1){13}}
\put(35,12){\vector(2,1){13}}
\end{picture}

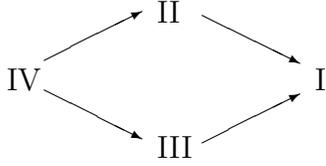
{\bf Figure 2.}  Blow-Down Diagram
\vspace{0.5cm}

Each of the fans $\Sigma_{\text{I--IV}}$ 
defines a coherent subset $S_{\text{I--IV}}$ of 
$L(P)^{(4)}$, the set of the codimension four faces of $P$.
Then according to (\ref{result}),
we can construct the maximal cones
$K_{\text{I--IV}}:=K(S_{\text{I--IV}})$ 
of the \Ka moduli space $\kahler^{12}(1,1,5,5)$.
The result is as follows~:
\begin{equation}
K_{\text{I}}=\pos 
\left\{
\begin{gathered}
-\ve_3+\ve_4-\ve_7+\ve_8-\ve_{11},\ 
-\ve_4+\ve_9,\
-\ve_8+\ve_9,\
-\ve_7    \\
-\ve_3+\ve_8,\ 
-\ve_3+\ve_4,\
-\ve_1+\ve_2,\
-\ve_4+\ve_5,\
-\ve_5+\ve_{10}\\
-\ve_{11},\ 
-\ve_7+\ve_9-\ve_{11},\ 
-\ve_3+\ve_6,\
-\ve_4+\ve_6-\ve_8+\ve_9 \\
-\ve_7+\ve_{10},\
-\ve_1+\ve_2-\ve_4+\ve_6,\
-\ve_5+\ve_6-\ve_8+\ve_{10}\\
-\ve_1+\ve_2-\ve_5+\ve_6-\ve_9+\ve_{10},\ 
-\ve_3+\ve_5-\ve_7+\ve_8\\
-\ve_1+\ve_6,\
-\ve_4+\ve_5-\ve_7+\ve_9
\end{gathered}
\right\}.
\end{equation}
\begin{equation}
K_{\text{II}}=\pos 
\left\{
\begin{gathered}
-\ve_7,\
-\ve_3+\ve_4-\ve_5+\ve_6-\ve_9+\ve_{10}-\ve_{11},\
-\ve_3+\ve_6 \\
-\ve_5+\ve_6-\ve_8+\ve_{10},\
-\ve_1+\ve_2-\ve_5+\ve_6-\ve_9+\ve_{10} \\
-\ve_3+\ve_8,\
-\ve_3+\ve_4,\
-\ve_5+\ve_{10},\
-\ve_9+\ve_{10},\
-\ve_1+\ve_6    \\
-\ve_{11},\
-\ve_3+\ve_4-\ve_7+\ve_8-\ve_{11}
\end{gathered}
\right\}.
\end{equation}
\begin{equation}
K_{\text{III}}=\pos 
\left\{
\begin{gathered}
-\ve_{11},\
-\ve_1+\ve_2,\
-\ve_1+\ve_2-\ve_3+\ve_6-\ve_7+\ve_8-\ve_9 \\
-\ve_1+\ve_2-\ve_4+\ve_6,\
-\ve_1+\ve_2-\ve_5+\ve_6-\ve_9+\ve_{10} \\
-\ve_4+\ve_5,\
-\ve_3+\ve_8,\
-\ve_3+\ve_4,\
-\ve_1+\ve_6,\
-\ve_3+\ve_6   \\
-\ve_7,\
-\ve_3+\ve_5-\ve_7+\ve_8,\
-\ve_3+\ve_4-\ve_7+\ve_8-\ve_{11}
\end{gathered}
\right\}.
\end{equation}
\begin{equation}
K_{\text{IV}}=\pos 
\left\{
\begin{gathered}
-\ve_{11},\
-\ve_3+\ve_8,\
-\ve_3+\ve_4-\ve_5+\ve_6-\ve_9+\ve_{10}-\ve_{11}\\
-\ve_7,\
-\ve_3+\ve_4,\
-\ve_1+\ve_2-\ve_3+\ve_6-\ve_7+\ve_8-\ve_9\\
-\ve_1+\ve_2-\ve_5+\ve_6-\ve_9+\ve_{10},\
-\ve_1+\ve_6,\
-\ve_3+\ve_6  \\
-\ve_9+\ve_{10},\
-\ve_3+\ve_4-\ve_7+\ve_8-\ve_{11}
\end{gathered}
\right\}.
\end{equation}

\subsubsection{1/5(1,2,3,4) model}
Let us first choose the following
representatives for the $\Ga$-orbits~: 
\begin{alignat}{4}
{\cal F}_{5}&={\cal F}_{5}^{(4)},  &\quad
{\cal F}_{6}&={\cal F}_{6}^{(2)},  &\quad
{\cal F}_{7}&={\cal F}_{7}^{(0)},  &\quad
{\cal F}_{8}&={\cal F}_{8}^{(0)}, \nonumber \\
{\cal F}_{9} &={\cal F}_{9}^{(1)},   &\quad
{\cal F}_{10}&={\cal F}_{10}^{(3)},  &\quad
{\cal F}_{11}&={\cal F}_{11}^{(0)},  &\quad
{\cal F}_{12}&={\cal F}_{12}^{(0)},
\label{choice-1}
\end{alignat}
which satisfy the compatibility condition~:
\begin{equation}
\bigcap_{k=5}^{12}\pi_{\Q}
\left({\cal F}_{k}\right)
=\pos
\{\skima 
-\ve_1,\ 
-\ve_2,\
-\ve_3,\
-\ve_4
\}
\label{zentaide}
\end{equation}
If we define the phase I by
\vspace{0.5cm}

phase I ($\Sigma_{\text{I}}$)
\begin{align}
&\langle 2,3,10,11  \rangle,  \quad
\langle 1,4,7,12  \rangle,   \quad
\langle 3,5,7,10  \rangle,   \quad
\langle 2,3,8,11  \rangle,   \quad
\langle 1,7,8,12  \rangle,   \nonumber \\
&\langle 2,6,8,11  \rangle,   \quad
\langle 2,3,4,5  \rangle,    \quad
\langle 1,3,7,8  \rangle,    \quad
\langle 2,4,5,6  \rangle,    \quad
\langle 1,2,6,8  \rangle,\nonumber \\
&\langle 1,4,6,9  \rangle,   \quad
\langle 1,2,4,6  \rangle,   \quad
\langle 1,3,4,7  \rangle,   \quad
\langle 1,2,3,8  \rangle,   \quad
\langle 1,4,9,12 \rangle,\nonumber \\
&\langle 3,4,5,7   \rangle,  \quad
\langle 2,3,5,10  \rangle,  \quad
\langle 4,5,6,9   \rangle,   \\
&\langle 3,7,8,10,11 \rangle,   \quad
\langle 2,5,6,10,11  \rangle,  \quad
\langle 4,5,7,9,12  \rangle,   \quad
\langle 1,6,8,9,12  \rangle,\nonumber \\
&\langle 5,6,7,8,9,10,11,12  \rangle, \nonumber
\label{hilb-1234-5}
\end{align}
the associated cone 
$K_{\text{I}}:=K(S_{\text{I}})$ 
coincides with (\ref{zentaide}).
Therefore the phase I is the only possible phase
under the choice of the representatives 
for the exceptional divisors (\ref{choice-1}).
\vspace{0.5cm}

The second choice of the compatible representatives
for the exceptional divisors is~:
\begin{alignat}{4}
{\cal F}_{5}&={\cal F}_{5}^{(4)}, &\quad
{\cal F}_{6}&={\cal F}_{6}^{(2)}, &\quad
{\cal F}_{7}&={\cal F}_{7}^{(0)}, &\quad
{\cal F}_{8}&={\cal F}_{8}^{(1)}, \nonumber \\
{\cal F}_{9} &={\cal F}_{9}^{(1)},  &\quad
{\cal F}_{10}&={\cal F}_{10}^{(3)}, &\quad
{\cal F}_{11}&={\cal F}_{11}^{(1)}. &\quad
\phantom{{\cal F}_{12}}&\phantom{={\cal F}_{12}^{(0)}.}
\label{choice-2}
\end{alignat}
Note the absence of the facet associated with
$\vwb_{12}$ in (\ref{choice-2}).

Consider the following two phases
\vspace{0.5cm}

phase II ($\Sigma_{\text{II}}$)
\begin{align}
&\langle 5,8,10,11   \rangle,\quad
\langle 2,3,4,5  \rangle,\quad
\langle 3,4,5,7   \rangle,\quad
\langle 4,5,6,8   \rangle, \nonumber \\
&\langle 1,2,3,8   \rangle,\quad
\langle 2,3,8,11   \rangle,\quad
\langle 1,4,5,8   \rangle,\quad
\langle 3,8,10,11   \rangle,\nonumber   \\
&\langle 2,5,10,11   \rangle,\quad
\langle 1,4,6,8   \rangle,\quad
\langle 1,3,4,7   \rangle,\quad
\langle 2,3,10,11   \rangle,  \\
&\langle 1,2,6,8   \rangle,\quad
\langle 2,3,5,10   \rangle,\quad
\langle 2,4,5,6   \rangle,\quad
\langle 3,5,7,10   \rangle, \nonumber   \\
&\langle 1,4,5,7   \rangle,\quad
\langle 1,2,4,6   \rangle, \nonumber \\
&\langle 1,3,7,8,10   \rangle,\quad
\langle 1,5,7,8,10   \rangle,\quad
\langle 2,5,6,8,11   \rangle.\nonumber
\end{align} 

phase III ($\Sigma_{\text{III}}$)
\begin{align}
&\langle 3,8,10,11   \rangle,\quad
\langle 3,5,7,10   \rangle,\quad
\langle 1,2,3,8   \rangle,\quad
\langle 2,3,4,5   \rangle,  \nonumber \\
&\langle 2,3,8,11   \rangle,\quad
\langle 1,2,6,8   \rangle,\quad
\langle 5,8,10,11   \rangle,\quad
\langle 2,3,10,11   \rangle, \nonumber \\
&\langle 2,5,10,11   \rangle,\quad
\langle 1,2,4,6   \rangle,\quad
\langle 3,4,5,7   \rangle,\quad
\langle 4,5,6,9   \rangle,  \\
&\langle 1,4,5,9   \rangle,\quad
\langle 2,3,5,10   \rangle,\quad
\langle 2,4,5,6   \rangle,\quad
\langle 1,3,4,7   \rangle,\nonumber  \\
&\langle 1,4,6,9   \rangle,\quad
\langle 1,4,5,7   \rangle, \nonumber \\
&\langle 2,5,6,8,11  \rangle,\quad
\langle 1,3,7,8,10   \rangle,\quad
\langle 1,5,7,8,10   \rangle,\quad
\langle 1,5,6,8,9   \rangle.\nonumber
\end{align}
Using (\ref{result}), we see that 
these two phases can be realized in the maximal cones
$K_{\text{II}}$ and $K_{\text{III}}$ defined by
\begin{gather}
K_{\text{II}}=\pos
\left\{
-\ve_3,\ 
\ve_1-\ve_3-\ve_4,\
\ve_1-\ve_2-\ve_3-\ve_4,\
\ve_1-\ve_2-\ve_3
\right\},\\
K_{\text{III}}=\pos
\left\{
-\ve_3,\ 
\ve_1-\ve_3-\ve_4,\
\ve_1-\ve_2-\ve_3-\ve_4,\
-\ve_4
\right\}.\phantom{-\ve_2-\ve_3}
\end{gather}
So far we have seen only the phases the fan of which 
is not simplicial, which means that the singularity of 
$\Mod$ is worse than orbifold ones in these phases.

In fact, combinatorics of the facets of $P$
admits neither 
the smooth phases incorporating 
two age 3 divisors, for example,
\begin{alignat}{5}
&\langle 3,7,8,10  \rangle,   &\quad
&\langle 5,6,8,11  \rangle,   &\quad
&\langle 3,8,10,11  \rangle,   &\quad
&\langle 2,5,10,11  \rangle,   &\quad
&\langle 2,3,4,5    \rangle,   \nonumber \\
&\langle 1,4,7,8  \rangle,   &\quad
&\langle 2,4,5,6  \rangle,   &\quad
&\langle 1,2,4,6  \rangle,   &\quad
&\langle 3,4,5,7  \rangle,   &\quad
&\langle 4,5,7,8  \rangle,   \nonumber \\
&\langle 5,8,10,11  \rangle,   &\quad
&\langle 5,7,8,10  \rangle,   &\quad
&\langle 4,5,6,8  \rangle,   &\quad
&\langle 2,5,6,11  \rangle,   &\quad
&\langle 2,3,8,11  \rangle,   \\
&\langle 1,4,6,8  \rangle,   &\quad
&\langle 2,6,8,11  \rangle,   &\quad
&\langle 1,3,4,7  \rangle,   &\quad
&\langle 2,3,10,11  \rangle,   &\quad
&\langle 2,3,5,10  \rangle,   \nonumber \\
&\langle 3,5,7,10  \rangle,   &\quad
&\langle 1,3,7,8  \rangle,   &\quad
&\langle 1,2,6,8  \rangle,   &\quad
&\langle 1,2,3,8  \rangle,   &\quad
&\phantom{\langle 1,1,1,1\rangle,}\nonumber
\end{alignat}
nor the phase with the simplicial fan incorporating
all the exceptional divisors~:
\begin{alignat}{5}
&\langle 1,4,7,12  \rangle,   &\quad
&\langle 1,4,9,12  \rangle,   &\quad
&\langle 8,9,10,11  \rangle,   &\quad
&\langle 5,6,9,11  \rangle,   &\quad
&\langle 4,7,9,12  \rangle,   \nonumber \\
&\langle 2,5,10,11  \rangle,   &\quad
&\langle 1,7,8,12  \rangle,   &\quad
&\langle 7,8,9,12  \rangle,   &\quad
&\langle 1,8,9,12  \rangle,   &\quad
&\langle 2,3,4,5  \rangle,   \nonumber \\
&\langle 1,3,4,5  \rangle,   &\quad
&\langle 1,3,4,7  \rangle,   &\quad
&\langle 1,2,4,6  \rangle,   &\quad
&\langle 3,4,5,7  \rangle,   &\quad
&\langle 2,3,8,11  \rangle,   \nonumber \\
&\langle 4,5,7,9  \rangle,   &\quad
&\langle 2,6,8,11  \rangle,   &\quad
&\langle 1,4,6,9  \rangle,   &\quad
&\langle 6,8,9,11  \rangle,   &\quad
&\langle 4,5,6,9  \rangle,   \\
&\langle 2,3,10,11  \rangle,   &\quad
&\langle 3,8,10,11  \rangle,   &\quad
&\langle 5,9,10,11  \rangle,   &\quad
&\langle 3,7,8,10  \rangle,   &\quad
&\langle 7,8,9,10  \rangle,   \nonumber \\
&\langle 5,7,9,10  \rangle,   &\quad
&\langle 2,5,6,11  \rangle,   &\quad
&\langle 2,3,5,10  \rangle,   &\quad
&\langle 3,5,7,10  \rangle,   &\quad
&\langle 1,6,8,9  \rangle,   \nonumber \\
&\langle 1,2,6,8  \rangle,   &\quad
&\langle 1,2,3,8  \rangle,   &\quad
&\langle 1,3,7,8  \rangle,   &\quad
&\phantom{\langle 1,1,1,1  \rangle,}   &\quad
&\phantom{\langle 1,1,1,1  \rangle,}\nonumber
\end{alignat}
the fans of which are found by the Delaunay triangulations 
\cite{Del}, \cite[p.~146]{ziegler} of $6T$.

\single
\section{$\boldsymbol\Ga$-Hilbert Scheme}
\renewcommand{\theenumi}{\roman{enumi}}
\renewcommand{\labelenumi}{(\theenumi)}

\subsection{Symplectic Quotient Construction}
%
%
Let $X$ be a quasi-projective variety with a fixed embedding
in a projective space.
The Hilbert scheme $\text{Hilb}^P(X)$ is 
the moduli space that parametrizes all the closed subschemes
of $X$ with a fixed Poincar\'e polynomial $P(z)$,
where $P(l)\in {\Z}$, for all $l\in {\Z}$.
See \cite{hilbert},\  \cite[Chapter I]{kollar}
for more detailed informations.

Let us take the following pair~:
$X=\C^d$, $P(z)=n$ (constant),
and consider the moduli space of
zero-dimensional closed subschemes
of length $n$ in $\C^d$,
which we denote by $\Hn$
\cite{ito-nakajima,ito-nakamura,
nakajima,reid}.
A point $Z\in \Hn$ corresponds 
to a ideal $I\subset A$
of colength $n$, where
$A=:{\C}[x_1,\dots ,x_d]$
is the coordinate ring of $\C^d$.
Therefore we have
\begin{equation}
\Hn=\left\{\skima \mathrm{ideal}\ I\subset A \left|\
\dim_{\C} A/I = n  \right. \right\}.     
\label{Hilbn}
\end{equation}
For $d,n \geq 3$, $\Hn$ is a singular variety.

The action of $\Ga$ on ${\C}^d$ 
is naturally extended to
that on $\Hn$.
Let $(\Hn)^{\Ga}$ be the subset of 
$\Hn$ which is fixed by the action of  
$\Ga$.
Each  point of $(\Hn)^{\Ga}$ corresponds
to a $\Ga$-invariant ideal $I$ of $A$. 
Consequently, for  $I\in (\Hn)^{\Ga}$,
$A/I=H^0(Z,{\cal O}_Z)$ becomes 
a $\Ga$-module of rank $n$.

For example, $\Ga$-orbit of a point 
$p\in \C^d\setminus\bold{0}$ 
is a point of $(\Hn)^{\Ga}$,  and it
constitutes the regular representation
$R$ (\ref{regular-rep}) of $\Ga$.

Now we give a definition of the $\Ga$-Hilbert scheme
following \cite{ito-nakajima}~:
\begin{equation}
\Hilb:=
\left.
\left\{
I\in (\Hn)^{\Ga} 
\right|\ 
 A/I\cong R\ 
\right\},
 \label{Gamma-Hilb}
\end{equation}
which means that the $\Ga$-Hilbert scheme $\Hilb$
parametrizes all the zero-dimensional closed 
subschemes $Z\subset {\C}^d$ such that
$H^0(Z,{\cal O}_Z)$ is isomorphic to the
regular representation $R$ of $\Ga$.
The mathematical aspect  
of the $\Ga$-Hilbert scheme $\Hilb$ 
for $d=2,3$ has been largely uncovered~:
For $d=2$, $\hilb^{\Ga}(\C^2)$ is a minimal resolution 
of the singularity $1/m(1,m-1)$ \cite{ito-nakamura}~;
What is more, it has been shown that even for $d=3$, 
$\hilb^{\Ga}(\C^3)$ is a crepant resolution 
of the Calabi--Yau three-fold singularity $\C^3/\Ga$
by I.~Nakamura in \cite{nakamura}, despite of the fact 
that $\hilb^n(\C^3)$ itself is {\it singular}.

Thus our interest here is also
concentrated on the $d=4$ case.
We will show later that $\hilb^{\Ga}(\C^4)$ 
is singular in general.

The definition of Hilbert schemes $\Hn$ and $\Hilb$ 
given above may seem abstract.
However Y.~Ito and H.~Nakajima has shown that
they can be realized as holomorphic (GIT)/symplectic 
quotients of flat spaces associated with the gauge group 
$\mathrm{U}(n)$ and $\mathrm{U}(1)^{n}$ 
respectively \cite{ito-nakajima,nakajima},
that is, we can identify $\Hn$ and $\Hilb$ 
as the classical Higgs moduli spaces 
of supersymmetric gauge theories \cite{MP,witten1}.
In particular $\Hilb$ can be described as a toric variety.
Furthermore, it is isomorphic to the D-brane \config\ space 
$\Mod$ with a particular choice 
of the Fayet--Iliopoulos parameter $\vr\in M^{'}_{\Q}$
\cite{ito-nakajima},
which is the main point of this subsection.

Let us first explain a holomorphic quotient construction of
$\Hn$. 
Fix $I\in \Hn$ and let $V=A/I$ be the associated
$n$ dimensional vector space.
The multiplication of $x_{\mu}$ on $V$ 
defines $d$-tuple of the elements of $\text{End}(V)$
which we denote by $X_{\mu}$.
To be more explicit, we choose an arbitrary basis  
$\vep_i$, $i=1,\dots ,n$ of $V$,
and we define the matrix elements of $X_{\mu}$ by
$(x_{\mu}+I)\cdot \vep_i=\sum_{j=1}^{n}(X_{\mu})^j_i\ \vep_j$.
If we also define a basis of $\C^d$ by $\vbeta_{\mu}$,
$\mu=1,\dots ,d$,
then we define an element $X$ 
of $\Hom(V,\C^d\otimes V)$ 
by
\begin{equation}
X(\vep_i):=
\sum_{\mu=1}^d \sum_{j=1}^n 
\vbeta_{\mu}\otimes \vep_j\ 
(X_{\mu})^j_i.
\end{equation}
Similarly the image of the map 
$1\hookrightarrow A\ra A/I$
defines a non-zero element of $V$ which we denote by
$p(1)=\sum_{i=1}^np^i\ \vep_i$,
where we mean by $p$
the associated element of $\Hom(\C,V)$, that is,\ 
$p(\la):=\la\ p(1)$\  for $\la \in \C$.

It is clear by construction that $(X,p)$ satisfies 
the following two conditions~:
\begin{enumerate}
\item $\left[X_{\mu}, X_{\nu}\right]=O$    \qquad
(F-flatness).
\item 
$p(1)$ is a cyclic vector, that is,
$V$ is generated by $X_{\mu}$ over $p(1)$ 
\ \ (stability). 
\end{enumerate}

Conversely let $\Vect$ be the vector space
$\Hom(\C^n,\C^d\otimes \C^n)\oplus \Hom(\C,\C^n)$,
and take such an element  $(X,p)\in \Vect$
that satisfies the above two conditions (i), (ii) with $V=\C^n$.
Then $(X,p)$ defines a point $I\in\Hn$ as follows~: 
First we define a surjective homomorphism 
$\kappa: A \ra {\C}^n$ 
of vector spaces over $\C$ by
\begin{equation}
\kappa(x_{\mu_1}\cdots x_{\mu_s}):=
X_{\mu_1}\cdots X_{\mu_s}\cdot p(1) 
=\sum_{i_1,\dots ,i_s,j}
\vep_{i_1}\skima (X_{\mu_1})^{i_1}_{i_2}\skima 
(X_{\mu_2})^{i_2}_{i_3}\skima \cdots  \skima
(X_{\mu_s})^{i_s}_{j}
\medspace p^j, 
\end{equation}
where we define
$X_{\mu}(\vep_i):=\sum_{j=1}^n \vep_j\ (X_{\mu})^j_i$,
and 
$p(1):=\sum_{i=1}^n p^i\skima \vep_i$ 
for a basis $\vep_i$ of $\C^n$~; 
Second  let  $I:=\Ker \kappa \subset A$
be an ideal of $A$,
then $A/I\cong {\C}^n$ as a vector space, 
which implies $I\in\Hn$ according to (\ref{Hilbn})~;
Third, it is clear that two elements 
 $(X,p)$ and  $(X^{'},p^{'})$ of $\Vect$
define the same point of $\Hn$
if and only if 
\begin{equation*}
(X^{'},p^{'})=(g X g^{-1},gp),\quad
\exists g\in \text{GL}(n,{\C}).
\end{equation*}
Thus we have arrived at the following
holomorphic quotient construction  of $\Hn$~:
\begin{equation}
\Hn\cong 
\left\{
(X,p) \in \Vect\ 
\left|\ 
\begin{aligned}
&\text{condition (i) : F-flatness}\\
&\text{condition (ii) : stability}
\end{aligned}
\right\}
\right/
\text{GL}(n,\C).
\end{equation}
We can also obtain 
the corresponding 
symplectic quotient construction of $\Hn$
by replacing the stability condition (ii) and
the quotient by $\text{GL}(n,\C)$ above
by the D-flatness condition
\begin{equation}
D_{\kahilb}:=\sum_{\mu=1}^{d}
\left[X_{\mu},X_{\mu}^{\dagger}\right]
+p\cdot p^{\dagger}
-\kahilb\skima \text{diag}(1,\dots ,1)
=O,
\label{D-flat}
\end{equation}
and the quotient by $\text{U}(n)$,
where $\kahilb>0$
is a unique Fayet--Illiopoulos parameter
associated with the $\text{U}(1)$ factor of
$\text{U}(n)$.

If we set $\kahilb=0$,
we obtain the symmetric product  
$(\C^d)^n/{\frak S}_n$ as a quotient variety 
reflecting the existence of the Hilbert-Chow morphism
$\Hn\ra (\C^d)^n/{\frak S}_n$.
\vspace{1cm}

Let us turn to the holomorphic quotient construction
of $\Hilb$ based on that of $\Hn$ given above.
The only difference {}from the previous treatment 
of $\Hn$ is that this time 
we must assign the $\Ga$-quantum numbers
to the objects : $x_{\mu}$, 
$\vep_i$, $\vbeta_{\mu}$ and $p$.
However we would not mind repeating 
almost the same argument for convenience.

Let us first redefine the generator $g$ of $\Ga$ 
so that the action of which on $x_{\mu}$ becomes
$g\cdot x_{\mu}= \omega^{-\amu}x_{\mu}$
for consistency.
Second take a point $I \in \Hilb$ and 
define  $V=A/I$, which is now isomorphic to
the regular representation
$R$ as a $\Ga$-module by definition.
Let $\vep_i \in V$ be a generator of
$R_i$ for $i=1,\dots ,n$, that is,
$g\cdot \vep_i=\omega^i \vep_i$ and 
$V=\bigoplus_{i=0}^{n-1}\C \ \vep_i$ 
is the irreducible decomposition of $\Ga$-modules.

We also introduce somewhat abstractly 
$\vbeta_{\mu}$
as a generator of $R_{\amu}$  
for $\mu=1,\dots ,d$ 
and let
$Q=\bigoplus_{\mu=1}^d \C\  \vbeta_{\mu}$ 
be a $\Ga$-module.

Then we define a $\Ga$-equivariant homomorphism
{}from $V$ to $Q\otimes V$, which we call $X$, by
\begin{equation}
X : f\in V \ra 
\sum_{\mu=1}^d 
\vbeta_{\mu}\otimes (f\cdot x_{\mu})\in Q\otimes V,
\end{equation}
where the product of polynomials $f\cdot x_{\mu}$
is evaluated modulo $I$.
In particular $\mu$~th component 
of $X(\vep_{i+\amu})$ becomes
\begin{equation}
x_{\mu}\cdot \vep_{i+\amu} =(X_{\mu})^{i}_{i+\amu} \ 
 \vep_{i}, \quad \exists (X_{\mu})^{i}_{i+\amu} \in \C.
\end{equation}
Thus we get the matrices $(X_{\mu})$ 
of the same content as
those for a D-brane at the orbifold singularity.
The map $1\hookrightarrow A\ra A/I$ now induces
an element $0\ne p\in \Hom_{\Ga}(\C,V)$, where
$p(1)=p^0\skima \vep_0\in V$.
Thus an element $I\in \Hilb$ defines an element
$(X,p)\in
\Hom_{\Ga}(V,Q\otimes V)
\oplus \Hom_{\Ga}(\C,V)$ 
and it clear  by construction that $(X,p)$ satisfies
the conditions (i) (F-flatness) and (ii) (stability) above.

Conversely 
take an element $(X,p)$ of
$\Vect^{\skima\Ga}:=\Hom_{\Ga}(R, Q\otimes R)
\oplus \Hom_{\Ga}(\C,R)$
such that $(n,n)$ matrices $(X_{\mu})$ 
and $(n,1)$ matrix $p(1)$ 
defined by
\begin{equation*}
X(\vep_i)=\sum_{\mu=1}^d 
(X_{\mu})^{i-\amu}_i\ 
\vbeta_{\mu}\otimes \vep_{i-\amu},\qquad 
p(1)=p^0 \ \vep_0,
\end{equation*}
satisfies the conditions (i), (ii) with $V=R$.
Then we can define a $\Ga$-equivariant
surjective homomorphism $\kappa :A\ra R$ by
\begin{equation}
\kappa(x_{\mu_1}\cdots x_{\mu_s})
:=X_{\mu_1}\cdots X_{\mu_s}\cdot p(1)
=\sum_{i_1,\dots ,i_s}
\skima \vep_{i_1} \skima (X_{\mu_1})^{i_1}_{i_2} 
\skima 
(X_{\mu_2})^{i_2}_{i_3}\skima \cdots \skima
(X_{\mu_s})^{i_s}_{0} \medspace p^0,
\end{equation}
so that the ideal 
$I:=\Ker \kappa$ 
is $\Ga$-invariant
and we obtain the $\Ga$-module isomorphism
$A/I\cong R$, that is, $(X,p)$ defines 
an element $I$ of $\Hilb$.
With the basis $\vbeta_{\mu}$ of $Q$ fixed, 
two elements 
$(X,p)$ and $(X^{'},p^{'})$ 
of
$\Vect^{\skima\Ga}$ 
which satisfy the conditions (i) and (ii)
define the same point $I\in \Hilb$ if and only if
they are related  as
$(X^{'},p^{'})=(uX u^{-1},up)$
by an element 
$u:=(u_i) \in \Aut_{\Ga}(R)
\cong \prod_{i=1}^n\Aut(R_i)
\cong (\C^*)^n$,
where $u_i$ acts on $\vep_i$ by
$\vep_i\ra u_i^{-1}\ \vep_i$.

Thus  we get the holomorphic quotient construction
of $\Hilb$~:
\begin{equation}
\Hilb\cong 
\left\{ 
(X,p)\in \Vect^{\skima\Ga}
\left|\ 
\begin{aligned}
&\text{condition (i) :  F-flatness}\\
&\text{condition (ii) :  stability}
\end{aligned}
\right\}
\right/
\prod_{i=1}^n\Aut(R_i),
\end{equation}
which in particular shows that $\Hilb$ is toric.
The associated symplectic quotient can be obtained  
by replacing the stability condition (ii) 
and the quotient by $\prod_{i}\Aut(R_i)$
by the D-flatness condition 
that takes the {\em same form} as (\ref{D-flat}) 
followed by the quotient 
by $\prod_{i}\text{U}(R_i)\cong \text{U}(1)^n$.
Consequently, the Fayet--Iliopoulos parameters 
associated with $\text{U}(1)^n$ is $\kahilb (1,\dots ,1)$.

To sum up, 
we have the symplectic quotient realization of $\Hilb$~: 
\begin{equation}
\Hilb\cong
\left\{
(X,p)\in \Vect^{\skima\Ga}
\left|\ 
\begin{aligned}
&\text{F-flatness :}\quad
\left[X_{\mu},X_{\nu}\right]=O \\
&\text{D-flatness :}
\quad  D_{\kahilb}=O 
\end{aligned}
\right\}
\right/
\text{U}(1)^n. 
\label{symplectic}
\end{equation}
The relation between $\Hilb$ and $\Mod$ can be easily seen
if we write down the D-flatness equations
for $\Hilb$ in components~:
\begin{equation}
\sum_{\mu=1}^{d}
\left(
|x_{\mu}^{(i)}|^2-|x_{\mu}^{(i-\amu)}|^2
\right)+|p^0|^2\  \delta^{i,0}
 \ = \ \kahilb, \quad i=0,1,\dots ,n-1,
\end{equation}
where we set
$x_{\mu}^{(i)}:=(X_{\mu})^{i}_{i+\amu}$ 
as before (\ref{as-before}).
We can delete $p^0$ and the diagonal $\text{U}(1)$
{}from the symplectic quotient construction
owing to the Higgs mechanism \cite{ito-nakajima}~: 
\begin{equation}
|p^0|^2=n\skima \kahilb.
\end{equation}
Then we are left with
the matrices $(X_{\mu})$,
which satisfy the same equations
as those of D-brane matrices 
(\ref{M-F-flat},\skima\ref{M-D-flat})
with the Fayet--Illiopulos parameter
\begin{equation}
(r_0,r_1,\dots ,r_{n-1})
=\kahilb\skima (-(n-1),1,\dots ,1)\in M^{'}_{\Q}.
\end{equation}
Thus we come to the conclusion~:
\begin{equation}
\Hilb\cong {\cal M}(\kahilb\skima\bold{1}),\quad
\bold{1}:=(\overbrace{1,\dots ,1}^{n-1}),
\label{conc1}
\end{equation}
where we have identified 
$M^{'}$ with $\Z^{n-1}$ by neglecting the zeroth component.
We also note the existence of 
the Hilbert-Chow morphism 
$\Hilb\ra \CY$,
which comes {}from the isomorphism~:
${\cal M}(\bold{0})\cong \CY$ \cite{sardo}.

\subsection{Another Algorithm for Computation}
The aim of this subsection is to translate 
the algorithm to compute the $\Ga$-Hilbert scheme 
given by Reid in \cite{reid}, which seems quite different
from the one given in the previous subsection,
into the language of convex polyhedra.
Closely related topics can be found in \cite{AGV,sturmfels}. 

Let $A=\C[x_1,\dots,x_d]$ the coordinate ring
of $\C^d$, where $g\in \Ga$ acts on $x_{\mu}$ as
multiplication by $\omega^{\amu}$, which defines 
the action of $\Ga$ on $\C^d$ from the {\it right}.
For $i=0,\dots, n-1$, we define ${\cal L}_i$ 
to be the ``orbifold line bundle'' on $\CY$
associated with the irreducible representation $R_{-i}$ 
of $\Ga$ where $g\in \Ga$ acts as multiplication 
by $\omega^{-i}$.
The global section of ${\cal L}_i$  is 
$\left(R_{-i}\otimes A\right)^{\Ga}$, that is,
the weight $i$ subspace of $A$.
Note that as a $\Ga$-module,
$A\cong \Sym Q=\oplus_{n=0}^{\infty}\skima S^n Q$.
The set of the monomial generators over $\C$ 
of $\left(R_{-i}\otimes A\right)^{\Ga}$,
which we denote by $M_i$, is given by
\begin{equation}
M_i=
\left\{ 
\vm\in ({\Z}_{\geq 0})^d 
\left| \  
\vm\cdot \va \equiv i \mod{n}
\right.
\right\},
\end{equation}
where $\vm =(m_1,\dots ,m_d)$ and $\va =(a_1,\dots ,a_d)$.
$M_0$ coincides with the coordinate ring 
of $\CY$, and each $M_i$ has a structure 
of a finitely generated $M_0$-module, 
the set of the generators of which
we denote by $B_i$.
Let $P_i=\conv  M_i$ 
be the {\it Newton polyhedron}
of the global monomial sections of ${\cal L}_i$,
which can be regarded as a polyhedron in $(\Mb_0)_{\Q}$,
where the lattice $\Mb_0$ is defined in (\ref{M-lattice}).
Then the toric variety $X(\Mb_0,P_i)$
defines the blow-up of $\CY=X(\Mb_0,P_0)$ 
by ${\cal L}_i$, 
which is denoted by $\text{Bl}_i(\CY)$. 
The normal fan ${\cal N}(P_i)$ in 
$(\Nb_0)_{\Q}$ (\ref{N-lattice})
is the fan associated with $\text{Bl}_i(\CY)$.
Evidently, $P_i$ can be 
expressed as the Minkowski sum
of the polytope $\conv  B_i$ 
and the cone $C_0=P_0$ (\ref{M-cone}).
On the other hand,
a celebrated theorem of E.~Noether adapted to 
$1/n(a_1,\dots ,a_d,n-i)$ model,
which is not Calabi--Yau, tells us
that all the members of $B_i$ can be found  
among those in $M_i$ of degree $\leq n$,
which implies the following way 
to construct the Newton polyhedron $P_i$ 
without any knowledge of $B_i$~: 
\begin{equation}
P_i\cong \conv  B_i^{'}+C_0,
\qquad
B^{'}_i:=
\big\{
\vm \in M_i
\big| \ 
\sum_{\mu=1}^d m_{\mu}\leq n
\big\}
\supset B_i.
\end{equation}
\begin{flushleft}
{\it Example.}
We take $1/5(1,2,3,4)$ model.
The four convex polyhedra are given by
\begin{align}
P_1&=\conv\skima \{\ve_1,\  3\ve_2,\  2\ve_3,\ 4\ve_4,\ 
 \ve_2+\ve_4\} +C_0,\nonumber \\
P_2&=\conv\skima \{2\ve_1, \ \ve_2, \ 4\ve_3,\ 
3\ve_4,\  \ve_3+\ve_4\}+C_0,\nonumber \\
P_3&=\conv\skima \{3\ve_1, \ 4\ve_2, \ \ve_3,\ 
2\ve_4, \ \ve_1+\ve_2\}+C_0,\nonumber \\
P_4&=\conv\skima \{4\ve_1, \ 2\ve_2, \ 3\ve_3,\
\ve_4, \ \ve_1+\ve_3\}+C_0.
\end{align}
\end{flushleft}
According to \cite{reid}, $\Hilb$ is the toric variety
associated with the fan in $(\Nb_0)_{\Q}$ 
that is the {\it coarsest common refinement} 
of the normal fans
${\cal N}(P_i)$, $i=1,\dots, n-1$,
which we denote by
${\cal N}( P_1)\ \cap\  \cdots \ \cap \ {\cal N}(P_{n-1})$.
To put differently, $\Hilb$ is the toric variety 
associated with the polyhedron $P_{\text{Hilb}}$ 
defined by
\begin{equation}
P_{\text{Hilb}}:=P_1+ \cdots + P_{n-1}
=\conv \left( B_1+\cdots +\ B_{n-1}\right)+C_0,
\label{conc2}
\end{equation}
because of the formula \cite[Proposition 7.12]{ziegler}~:
\begin{equation}
{\cal N}( P_1)\ \cap\  \cdots \ \cap \ {\cal N}(P_{n-1}) 
={\cal N}( P_1+\cdots + P_{n-1}). \label{special}
\end{equation}
It is clear by construction that 
$\Hilb$ is projective over $\CY=X(\Mb_0,C_0)$,
and that each $P_i$ defines a line bundle 
generated by global sections,
and $P_{\text{Hilb}}$ an ample one on $\Hilb$.

Note that $P_{\text{Hilb}}$ defined in (\ref{special})
is by no means a unique candidate for a polyhedron 
yielding the $\Ga$-Hilbert scheme~:
indeed any polyhedron of the form 
$\sum_{i=1}^{n-1}k_i\skima P_i$, where $k_i >0$,
for all $i$ fits for the job.
A distinguished feature of $P_{\text{Hilb}}$ (\ref{special})
among the family $\sum_{i=1}^{n-1}k_i\skima  P_i$  
is the following~:
 
\begin{flushleft}
{\it Conjecture.} 
{\sl Two polyhedra $(\Mb,Q(\widehat{\bold{1}}))$ 
and $(\Mb_0,P_{\mathrm{Hilb}})$ are isomorphic to each other 
modulo translation.}
\end{flushleft}
Recall that
$\widehat{\bold{1}}$ is an element of $M_{\Q}$ which satisfies
$\piQ(\widehat{\bold{1}})=\bold{1}$.

\single
\subsection{Computations}
Here we compute the $\Ga$-Hilbert schemes of some 
Calabi--Yau four-fold models to show 
the {\em power} of the formula (\ref{GIT=r}) 
of the toric quotient combined with (\ref{conc1}).
Another method (\ref{conc2}), though less effective,
serves as a consistency check 
of the result of (\ref{conc1}).

\begin{flushleft}
\subsubsection{(1/17)(1,1,6,9) model}
\end{flushleft}

The fan of the $\Ga$-Hilbert scheme is given by the following
collection of the maximal cones~:
\begin{alignat}{5}
&\langle 2,3,4,5 \rangle, &\quad 
&\langle 1,3,5,6 \rangle, &\quad 
&\langle 1,2,7,9 \rangle, &\quad 
&\langle 1,2,3,6 \rangle, &\quad 
&\langle 2,5,6,8 \rangle,         \nonumber \\
&\langle 1,2,4,7 \rangle, &\quad 
&\langle 2,3,5,6 \rangle, &\quad 
&\langle 1,2,8,9 \rangle, &\quad 
&\langle 1,7,8,9 \rangle, &\quad 
&\langle 2,7,8,9 \rangle,         \nonumber \\
&\langle 2,5,7,8 \rangle, &\quad 
&\langle 2,4,5,7 \rangle, &\quad 
&\langle 1,2,6,8 \rangle, &\quad 
&\langle 1,4,5,7 \rangle, &\quad 
&\langle 1,3,4,5 \rangle,         \nonumber \\
&\langle 1,5,6,8 \rangle, &\quad 
&\langle 1,5,7,8 \rangle, &\quad 
&\phantom{\langle 2,3,4,5 \rangle,} &\quad 
&\phantom{\langle 2,3,4,5 \rangle,} &\quad 
&\phantom{\langle 2,3,4,5 \rangle,}
\label{hilb-1169}
\end{alignat} 
where the weight vectors are
\begin{alignat}{3}
&\vwb_5=( 1, 1, 6, 9), &\quad
&\vwb_6=( 2, 1,12, 1), &\quad
&\vwb_7=( 3, 3, 1,10), \nonumber \\
&\vwb_8=( 4, 4, 7, 2), &\quad
&\vwb_9=( 6, 6, 2, 3). &\quad
&\phantom{\vwb_1=(17, 0, 0, 0)}
\end{alignat}
The fan (\ref{hilb-1169}) defines
one of the five crepant resolutions of
$(1/17)(1,1,6,9)$ model.

For other Calabi--Yau four-fold models
which admit crepant resolutions,
we only give the following conjecture.
\vspace{0.3cm}

\begin{flushleft}
{\it Conjecture.}
{\sl The $\Ga$-Hilbert schemes of
$1/(3m+1)(1,1,1,3m-2)$ and $1/(4m)(1,1,2m-1,2m-1)$ models
(\ref{trivial}) are the crepant resolutions of 
the corresponding orbifolds described in \cite{mohri}.}
\end{flushleft}

\subsubsection{1/5(1,2,3,4) model}
The $\Ga$-Hilbert scheme coincides with the phase I
found in the previous section (\ref{hilb-1234-5}).
\subsubsection{1/7(1,2,5,6) model}
The exceptional divisors appearing 
in the $\Ga$-Hilbert scheme are as follows~:
\begin{alignat}{4}
\vwb_5&=(1,2,5,6), &\quad
\vwb_6&=(2,4,3,5), &\quad
\vwb_7&=(3,6,1,4), &\quad
\vwb_8&=(4,1,6,3), \nonumber \\
\vwb_9&=(5,3,4,2), &\quad
\vwb_{10}&=(6,5,2,1), &\quad
\vwb_{11}&=(2,4,10,5), &\quad
\vwb_{12}&=(3,6,8,4), \nonumber \\
\vwb_{13}&=(4,8,6,3), &\quad
\vwb_{14}&=(5,3,4,9), &\quad
\vwb_{15}&=(5,10,4,2), &\quad
\vwb_{16}&=(6,5,2,8),  \\
\vwb_{17}&=(8,2,5,6), &\quad
\vwb_{18}&=(9,4,3,5), &\quad
\vwb_{19}&=(9,4,3,12), &\quad
\vwb_{20}&=(12,3,4,9).\nonumber
\end{alignat}
The fan of the $\Ga$-Hilbert scheme is given by
\begin{align} 
&\langle 1,2,3,10   \rangle, \quad
\langle 2,4,6,7      \rangle, \quad
\langle 1,2,7,10      \rangle, \quad
\langle 2,3,13,15      \rangle, \quad
\langle 2,6,12,13      \rangle, \nonumber \\
&\langle 2,3,12,13      \rangle, \quad
\langle 3,4,5,8      \rangle, \quad
\langle 2,3,10,15      \rangle, \quad
\langle 1,2,4,7      \rangle, \quad
\langle 2,3,11,12      \rangle, \nonumber \\
&\langle 2,3,5,11      \rangle, \quad
\langle 2,3,4,5     \rangle, \quad
\langle 1,3,8,9      \rangle, \quad
\langle 1,9,10,18      \rangle, \quad
\langle 4,5,6,14      \rangle, \nonumber \\
&\langle 1,4,19,20      \rangle, \quad
\langle 1,3,9,10      \rangle, \quad
\langle 2,4,5,6      \rangle, \quad
\langle 3,9,12,13      \rangle, \quad
\langle 6,9,12,13      \rangle,  \\
&\langle 1,4,17,20      \rangle, \quad
\langle 1,3,4,8      \rangle, \quad
\langle 1,4,16,19      \rangle, \quad
\langle 1,4,8,17      \rangle, \quad
\langle 1,4,7,16      \rangle, \nonumber \\
&\langle 2,7,10,15       \rangle, \quad
\langle 4,6,7,16      \rangle, \quad
\langle 3,5,8,11      \rangle, \quad
\langle 1,8,9,17  \rangle, \nonumber \\
&\langle 3,8,9,11,12      \rangle, \quad
\langle 2,6,7,13,15      \rangle, \quad
\langle 1,9,17,18,20      \rangle, \quad
\langle 4,6,14,16,19      \rangle, \nonumber \\
&\langle 3,9,10,13,15  \rangle, \quad
\langle 2,5,6,11,12 \rangle, \quad
\langle 1,16,18,19,20 \rangle, \quad
\langle 4,14,17,19,20  \rangle, \nonumber \\
&\langle 1,7,10,16,18      \rangle, \quad
\langle 4,5,8,14,17   \rangle, \nonumber \\
&\langle 6,7,9,10,13,15,16,18 \rangle, \quad
\langle 4,5,8,9,11,12,14,17  \rangle, \quad
\langle 6,9,14,16,17,18,19,20 \rangle.\nonumber
\end{align}
\subsubsection{1/7(1,1,2,3) model}
This model has seven weight vectors~:
\begin{alignat}{4}
\vwb_{5} &=(1,1,2,3),&\
\vwb_{6} &=(3,3,6,2),&\
\vwb_{7} &=(4,4,1,5),&\
\vwb_{8} &=(5,5,3,1),\\
\vwb_{9} &=(6,6,5,4),&\
\vwb_{10} &=(8,8,2,3), &\
\vwb_{11} &=(9,9,4,6).&\
\phantom{\vwb_{13}} &\phantom{=(2,2,2,2),}\nonumber
\end{alignat}
The fan of the $\Ga$-Hilbert scheme,
which is a smooth non-Calabi--Yau four-fold,
has only five of them~:
\begin{alignat}{5}
&\langle 2,4,5,7 \rangle, &\quad
&\langle 1,2,7,10 \rangle, &\quad
&\langle 1,4,5,7 \rangle, &\quad
&\langle 1,2,3,8 \rangle, &\quad
&\langle 1,4,5,8 \rangle, \nonumber \\
&\langle 2,4,5,8 \rangle, &\quad
&\langle 1,3,5,6 \rangle, &\quad
&\langle 2,3,5,6 \rangle, &\quad
&\langle 1,2,4,7 \rangle, &\quad
&\langle 1,3,4,5 \rangle, \\
&\langle 1,5,7,10 \rangle, &\quad
&\langle 1,2,8,10 \rangle, &\quad
&\langle 2,3,6,8 \rangle, &\quad
&\langle 1,5,8,10 \rangle, &\quad
&\langle 2,5,8,10 \rangle, \nonumber \\
&\langle 1,3,6,8 \rangle, &\quad
&\langle 2,3,4,5 \rangle, &\quad
&\langle 2,5,7,10 \rangle. &\quad
&\phantom{\langle 1,1,1,1 \rangle,} &\quad
&\phantom{\langle 2,2,2,2 \rangle,} \nonumber
\end{alignat}

\subsubsection{(1/16)(1,3,5,7) model}
The weight vectors appearing in the $\Ga$-Hilbert scheme 
are given by
\begin{alignat}{3}
\vwb_5&=(17,3,21,7), &\quad
\vwb_6&=(18,6,10,14), &\quad
\vwb_7&=(14,42,6,18), \nonumber \\
\vwb_8&=(5,15,9,3), &\quad
\vwb_9&=(11,33,7,13), &\quad
\vwb_{10}&=(12,4,12,20), \nonumber \\
\vwb_{11}&=(12,4,12,4), &\quad
\vwb_{12}&=(8,24,8,8), &\quad
\vwb_{13}&=(20,12,4,12), \nonumber \\
\vwb_{14}&=(13,7,33,11), &\quad
\vwb_{15}&=(3,9,15,5), &\quad
\vwb_{16}&=(6,2,14,10), \nonumber \\
\vwb_{17}&=(8,8,24,8), &\quad
\vwb_{18}&=(7,5,3,1), &\quad
\vwb_{19}&=(22,2,14,10),  \\
\vwb_{20}&=(18,6,42,14), &\quad
\vwb_{21}&=(30,10,6,18), &\quad
\vwb_{22}&=(18,6,10,30), \nonumber \\
\vwb_{23}&=(17,3,5,7), &\quad
\vwb_{24}&=(1,3,5,7), &\quad
\vwb_{25}&=(7,21,3,17), \nonumber \\
\vwb_{26}&=(13,7,1,11), &\quad
\vwb_{27}&=(23,5,3,17), &\quad
\vwb_{28}&=(10,14,2,6), \nonumber \\
\vwb_{29}&=(14,10,6,18), &\quad
\vwb_{30}&=(7,5,3,17), &\quad
\vwb_{31}&=(11,1,7,13), \nonumber \\
\vwb_{32}&=(4,12,4,12), &\quad
\vwb_{33}&=(17,3,5,23), &\quad
\vwb_{34}&=(10,14,2,22).\nonumber
\end{alignat}
The fan of the $\Ga$-Hilbert scheme is 
defined by the following 104 maximal cones~:
\begin{align}
&\langle 1,3,11,18 \rangle,  \quad
\langle 1,4,31,33 \rangle,  \quad
\langle 2,4,24,32 \rangle,  \quad
\langle 1,18,21,23 \rangle, \quad
\langle 2,9,12,32 \rangle,  \nonumber \\
&\langle 3,11,17,18 \rangle,   \quad
\langle 3,4,16,24  \rangle,   \quad
\langle 18,24,29,32 \rangle,  \quad
\langle 6,18,24,29  \rangle,  \quad
\langle 1,13,18,21  \rangle,  \nonumber \\
&\langle 1,11,18,23 \rangle, \quad
\langle 2,3,8,18   \rangle, \quad
\langle 2,25,28,34 \rangle, \quad
\langle 3,8,15,18  \rangle, \quad
\langle 2,26,28,34 \rangle, \nonumber \\
&\langle 2,4,26,34  \rangle, \quad
\langle 2,3,15,24  \rangle, \quad
\langle 2,8,12,18 \rangle, \quad
\langle 2,3,4,24  \rangle, \quad
\langle 2,8,15,24 \rangle, \nonumber \\
&\langle 6,11,16,24 \rangle, \quad
\langle 2, 9,12,18 \rangle, \quad
\langle  2,8,12,24 \rangle, \quad
\langle  8,15,18,24 \rangle, \quad
\langle  3,14,17,24 \rangle, \nonumber \\
&\langle 4,22,30,33  \rangle, \quad
\langle 11,17,18,24 \rangle, \quad
\langle 3,15,17,24 \rangle, \quad
\langle 4,26,30,34 \rangle, \quad
\langle 15,17,18,24 \rangle, \nonumber \\
&\langle 1,2,4,26   \rangle, \quad
\langle 1,21,23,27 \rangle, \quad
\langle 4,10,22,24 \rangle, \quad
\langle 6,10,22,24 \rangle, \quad
\langle 4,10,16,24 \rangle, \nonumber \\
&\langle 6,11,18,24 \rangle, \quad
\langle 6,10,16,24 \rangle, \quad
\langle 18,28,29,32 \rangle, \quad
\langle 1,4,26,27 \rangle, \quad
\langle 1,3,5,11 \rangle,  \nonumber \\
&\langle 1,2,26,28  \rangle, \quad
\langle 2,7,25,28 \rangle, \quad
\langle 1,13,26,28 \rangle, \quad
\langle 1,2,18,28 \rangle, \quad
\langle 1,13,18,28 \rangle, \nonumber \\
&\langle 13,18,28,29 \rangle, \quad
\langle 6,11,18,23 \rangle, \quad
\langle 6,23,29,30 \rangle, \quad
\langle 8,12,18,24  \rangle, \quad
\langle 13,18,21,29 \rangle, \nonumber \\
&\langle 1,3,5,19 \rangle, \quad
\langle 2,3,8,15 \rangle, \quad
\langle 3,11,14,17 \rangle, \quad
\langle 6,18,23,29 \rangle, \quad
\langle 1,2,3,18   \rangle, \nonumber \\
&\langle 2,4,25,34 \rangle, \quad
\langle 18,21,23,29 \rangle, \quad
\langle 3,5,16,19 \rangle, \quad
\langle 4,22,24,30 \rangle, \quad
\langle 6,22,24,30 \rangle, \nonumber \\
&\langle 4,26,27,30 \rangle, \quad
\langle 1,19,23,31 \rangle, \quad
\langle 1,3,4,31 \rangle, \quad
\langle 1,3,19,31 \rangle, \quad
\langle 2,7,9,32 \rangle, \nonumber \\
&\langle 3,5,11,20 \rangle, \quad
\langle 3,16,19,31 \rangle, \quad
\langle 3,4,16,31 \rangle, \quad
\langle 6,24,29,30 \rangle, \quad
\langle 4,10,16,31 \rangle, \nonumber \\
&\langle 9,12,18,32 \rangle, \quad
\langle 3,11,14,20 \rangle, \quad
\langle 2,7,25,32 \rangle, \quad
\langle 12,18,24,32 \rangle, \quad
\langle 2,12,24,32 \rangle, \nonumber \\
&\langle 2,4,25,32 \rangle, \quad
\langle 7,25,28,32 \rangle, \quad
\langle 1,23,27,33 \rangle, \quad
\langle 24,29,30,32 \rangle, \quad
\langle 4,24,30,32 \rangle, \nonumber \\
&\langle 1,4,27,33 \rangle, \quad
\langle 3,15,17,18 \rangle, \quad
\langle 23,27,30,33 \rangle, \quad
\langle 4,27,30,33 \rangle, \quad
\langle 1,23,31,33 \rangle, \nonumber \\
%
%
&\langle 3,5,16,20  \rangle, \quad
\langle 11,14,17,24 \rangle, \quad
\langle 5,11,16,20 \rangle, \nonumber  \\
&\langle 1,13,21,26,27 \rangle, \quad
\langle 2,7,9,18,28   \rangle, \quad
\langle 21,23,27,29,30 \rangle, \quad
\langle 6,22,23,30,33 \rangle, \nonumber \\
&\langle 4,10,22,31,33 \rangle, \quad
\langle 7,9,18,28,32 \rangle, \quad
\langle 1,5,11,19,23 \rangle, \quad
\langle 4,25,30,32,34 \rangle, \nonumber \\
&\langle 3,14,16,20,24 \rangle, \quad
\langle 11,14,16,20,24 \rangle,    \nonumber \\
&\langle 6,10,22,23,31,33 \rangle, \quad
\langle 5,6,11,16,19,23   \rangle, \quad
\langle 13,21,26,27,29,30   \rangle,\nonumber \\
&\langle 25,28,29,30,32,34 \rangle, \quad
\langle 13,26,28,29,30,34   \rangle, \quad
\langle  6,10,16,19,23,31  \rangle.   
\end{align}
We see that in general 
the $\Ga$-Hilbert scheme of a Calabi--Yau orbifold 
for $d=4$ is neither smooth nor Calabi--Yau
in contrast with the cases of $d=2,3$.
\vspace{0.5cm}

\begin{flushleft}
{\it Acknowledgement.}

I would like to thank Mitsuko Abe (Tokyo Inst. of Technology) 
for many discussions.
\end{flushleft}

\single


\begin{thebibliography}{99}
\bibitem{AGV}V.I.~Arnold, S.M.~Gusein-Zade and A.N.~Varchenko,
{\it Singularities of Differentiable Maps II},
 Monographs in Mathematics {\bf 82}, 
Birkh\"auser, Boston (1988).
\bibitem{AGM}P.S.~Aspinwall, B.R.~Greene and D.R.~Morrison,
Calabi--Yau Moduli Space, Mirror Manifolds 
and Spacetime Topology Change in String Theory,
{\it Nucl. Phys.} {\bf B416} (1994) pp.~414--480,
hep-th/9309097.
\bibitem{BD}V.V.~Batyrev and D.I.~Dais,
Strong McKay Correspondence, String Theoretic Hodge Numbers
and Mirror Symmetry,
{\it Topology} {\bf 35} (1996) pp.~901--929,
alg-geom/9410001.
\bibitem{BBMOOY}K.~Becker, M.~Becker, D.R.~Morrison,
H.~Ooguri, Y.~Oz and Z.~Yin,
Supersymmetric Cycles in Exceptional Holonomy Manifolds 
and Calabi--Yau Four-Folds,
{\it Nucl. Phys.} {\bf B480} (1996) pp.~225--238,
hep-th/9608116.
\bibitem{BVS}M.~Bershadsky, C.~Vafa and V.~Sadov,
D-Branes and Topological Field Theories,
{\it Nucl. Phys.} {\bf 463B} (1996) pp.~420--434,
hep-th/9511222. 
\bibitem{BFS}L.J.~Billera, P.~Fillman and B.~Sturmfels,
Constructions and Complexity of Secondary Polytopes,
{\it Adv. Math.} {\bf 83} (1990) pp.~155--179. 
\bibitem{cox}D.A.~Cox,
The Homogeneous Coordinate Ring of a Toric Variety,
{\it J. Alg. Geom.} {\bf 4} (1995) pp.~17--50,
alg-geom/9210008~; 
D.A.~Cox, 
Recent Developments in Toric Geometry,
alg-geom/9606016.
\bibitem{DHH}D.I.~Dais, U.-U.~Hause and M.~Henk,
On Crepant Resolutions of 2-Parameter Series 
of Gorenstein Cyclic Quotient Singularities, 
math/9803096.
\bibitem{DH} D.I.~Dais and M.~Henk,
On a Series of Gorenstein Cyclic Quotient Singularities
Admitting a Unique Projective Crepant Resolutions, 
math/9803094,
to appear in 
{\it Combinatorial Convex Geometry and Toric Varieties},
eds. G.M.~Ewald and B.~Teissier, 
Birkh\"auser, Boston.
\bibitem{DG}M.R.~Douglas and B.R.~Greene,
Metrics on D-Brane Orbifolds,
{\it Adv. Theor. Math. Phys.} {\bf 1} (1998) pp.~184--196,
hep-th/9707214.
\bibitem{DGM} M.R.~Douglas, B.R.~Greene and D.R.~Morrison,
{Orbifold Resolutions by D-Branes},
{\it Nucl. Phys.} {\bf B506} (1997) pp.~84--106,
hep-th/9704151.
\bibitem{DKPS}M.R.~Douglas, D.~Kabat,
 P.~Pouliot and S.H.~Shenker,
D-Brane and Short Distances in String Theory,
{\it Nucl. Phys.} {\bf B485} (1996) pp.~85--127,
hep-th/9608024.
\bibitem{DM}M.R.~Douglas and G.~Moore,
D-Branes, Quivers and ALE Instantons, \\
hep-th/9603167.
\bibitem{Del}H.~Edelsbrunner,
{\it Algorithms in Computational Geometry},
EATCS Monographs in Theoretical Computer Science 
{\bf 10}, Springer-Verlag, Berlin (1987).
\bibitem{ewald} G.M.~Ewald, 
{\it Combinatoric Convexity and Algebraic Geometry},
Graduate Texts in Mathematics {\bf 168},
Springer-Verlag, New York (1996).
\bibitem{fulton}W.~Fulton,
{\it Introduction to Toric Varieties},
Annals of Mathematics Studies {\bf 131},
Princeton Univ. Press, Princeton (1993).
\bibitem{GKZ}I.M.~Gelfand, M.M.~Kapranov 
and A.V.~Zelevinski,
{\it Discriminants, Resultants
 and Multidimensional Determinants}, 
Mathematics: Theory {\it\&} Applications,
Birkh\"auser, Boston (1994).
\bibitem{greene}B.R.~Greene,
D-Brane Topology Changing Transitions,
hep-th/9711124, to appear in {\it Nucl. Phys.} {\bf B}.
\bibitem{hilbert} A.~Grothendieck,
{\it Fondements de la G\'{e}om\'{e}trie Alg\'{e}brique},
Extraits du {\it S\'{e}minaire Bourbaki} 1957--1962, 
 mimeographed notes,
Secr\'{e}tariat math\'{e}matique,
 11 rue Pierre Curie, Paris 5e (1962).
\bibitem{hartshorne} R.~Hartshorne, 
{\it Algebraic Geometry},
Graduate Texts in Mathematics {\bf 52},
Springer-Verlag, New York (1977).
\bibitem{HLY}  S.~Hosono, B.H.~Lian and S.-T.~Yau,
GKZ-Generalized Hypergeometric Systems in Mirror Symmetry 
of Calabi--Yau Hypersurfaces,
{\it Comm. Math. Phys.} {\bf 182} (1996) pp.~535--577,
\ alg-geom/9511001.
\bibitem{ito-nakajima} Y.~Ito and H.~Nakajima,
{McKay Correspondence 
and Hilbert Schemes in Dimension Three},  
math/9803120.
\bibitem{ito-nakamura}
Y. Ito and I. Nakamura,
McKay Correspondence and Hilbert Schemes,
{\it Proc. Japan  Acad.} {\bf 72A} (1996) pp.~135--137~; 
Y. Ito and I. Nakamura,
Hilbert Schemes and Simple Singularities $A_n$ and $D_n$,
Hokkaido Univ. preprint {\bf 348} (1996)~;
I. Nakamura,
Hilbert Schemes and Simple Singularities 
$E_6$, $E_7$ and $E_8$,
Hokkaido Univ. preprint  {\bf 362} (1996).
\bibitem{KSZ}M.M.~Kapranov, B.~Sturmfels 
and A.V.~Zelevinsky,
Quotients of Toric Varieties,
{\it Math. Ann.} {\bf 290} (1991) pp.~643--655.
\bibitem{kollar}J.~Koll\'ar,
{\it Rational Curves on Algebraic Varieties},
Ergebnisse der Mathematik und ihrer Grenzgebiete, 
3. Folge {\bf 32},
Springer-Verlag, Berlin (1996).
\bibitem{kronheimer}
P.B.~Kronheimer, 
{The Construction of ALE Spaces 
as Hyper-K\"{a}hler Quotients},
{\it J. Diff. Geom.} {\bf 28} (1989) pp.~665--683.
\bibitem{mohri} K.~Mohri, 
{D-Branes and Quotient Singularities 
of Calabi--Yau Four-Folds},
{\it Nucl. Phys.} {\bf B521} (1998) pp.~161--182,
hep-th/9707012.
\bibitem{MP}D.R.~Morrison and M.R.~Plesser,
Summing the Instantons~: 
Quantum Cohomology and Mirror Symmetry in Toric Varieties,
{\it Nucl. Phys.} {\bf B440} (1995) pp.~279--354,
hep-th/9412236.
\bibitem{MS} D.R.~Morrison and G.~Stevens,
Terminal Quotient Singularities in Dimensions Three and Four,
{\it Proc. Amer. Math. Soc.} {\bf 90} (1984) pp.~15--20.
\bibitem{MR}S.~Mukhopadhyay and K.~Ray,
Conifolds from D-Branes, 
{\it Phys. Lett.} {\bf B423} (1998) pp.~247--254,
hep-th/9711131.
\bibitem{muto}T.~Muto, 
D-Branes on Orbifolds and Topology Change, 
{\it Nucl. Phys.} {\bf B521} (1998) pp.~183--201, 
hep-th/9711090.
\bibitem{nakajima} H.~Nakajima,
Lectures on Hilbert Schemes of Points on Surfaces,
preprint (1996), available via \     
http:$/\!/$www.kusm.kyoto-u.ac.jp/\~{}nakajima/TeX.html.
\bibitem{nakamura} I.~Nakamura, 
Hilbert Schemes of Abelian Group Orbits, 
preprint (1998), cited in \cite{ito-nakajima}. 
\bibitem{oda}T.~Oda, {\it Convex Bodies 
and Algebraic Geometry~:\  An Introduction 
to the Theory of Toric Varieties},
Ergebnisse der Mathematik und ihrer Grenzgebiete, 
3. Folge {\bf 15},
Springer-Verlag, Berlin (1988).
\bibitem{OOY}H.~Ooguri, Y.~Oz and Z.~Yin,
D-Branes on Calabi--Yau Spaces and Their Mirrors,
{\it Nucl. Phys.} {\bf B477} (1996)  pp.~407--430,
hep-th/9606112. 
\bibitem{ray}K.~Ray,
A Ricci-Flat Metric on D-Brane Orbifolds, 
hep-th/9803192.
\bibitem{reid}M.~Reid, 
McKay Correspondence, alg-geom/9702016. 
\bibitem{sardo} A.V.~Sardo~Infirri, 
{Partial Resolutions of Orbifold Singularities
 via Moduli Space of HYM-type Bundles},
alg-geom/9610004.
\bibitem{infirri}A.V.~Sardo~Infirri,
{Resolutions of Orbifold Singularities 
and the Transportation Problem
on the McKay Quiver}, 
alg-geom/9610005.
\bibitem{sturmfels} B.~Sturmfels,
{\it Gr\"obner Bases and Convex Polytopes},
University Lecture Series {\bf 8}, 
Americal Mathematical Society, Providence (1995).
\bibitem{thaddeus}M.~Thaddeus,
{Toric Quotients and Flips,}  in
 {\it Topology, Geometry and Field Theory}, 
eds K.~Fukaya, M.~Furuta, T.~Kohno and D.~Kotschick, 
World Scientific, Singapore (1994) pp.~193--213.
\bibitem{witten1} E.~Witten,
Phases of $N\!=\!2$ Theories in Two Dimensions,
{\it Nucl. Phys.} {\bf B403} (1993) pp.~159--222, 
hep-th/9301042.
\bibitem{witten2}E.~Witten,
Bound States of Strings and $p$-Branes,
{\it Nucl. Phys.} {\bf B460} (1996) pp.~335--350,
hep-th/9510135.
\bibitem{ziegler}G.~Ziegler, 
{\it Lectures on Polytopes},
Graduate Texts in Mathematics {\bf 152}, 
Springer-Verlag, New York (1994).
\end{thebibliography}
\end{document}